\documentclass[a4paper,12pt]{article}

\pdfoutput=1 

\usepackage{jcappub} 

\usepackage{latexsym,array,theorem,mathrsfs,subfigure,bm,float}
\usepackage{amsmath}
\usepackage{amsfonts}
\usepackage{amssymb}

\usepackage[compatibility=false]{caption}
\usepackage{epstopdf}
\usepackage{adjustbox}

\newcommand{\be}{\begin{eqnarray}}
\newcommand{\ee}{\end{eqnarray}}
\newcommand{\nn}{\nonumber}
\newcommand{\gm}{\gamma}
\newcommand{\mh}{m_{\rm H}}

\newcommand{\DAlambert}{\Box}
\newcommand{\gam}{\Gamma}

\newcommand{\gmm}{{\bm \gamma}}

\newcommand{\f}{\phi}
\newcommand{\h}{{h}}
\newcommand{\hh}{{\bf h}}

\newcommand{\M}{{\rm M}}

\newcommand{\e}{{\bm e}}

\newcommand{\muu}{M^2}

\renewcommand{\theequation}{\arabic{section}.\arabic{equation}}

\begin{document}

\title{Massive spin-2 field in arbitrary spacetimes  -- the detailed derivation}

\author[a]{Charles Mazuet,}
\emailAdd{\tt charles.mazuet@lmpt.univ-tours.fr}

\author[a,b]{Mikhail~S.~Volkov}
\emailAdd{\tt volkov@lmpt.univ-tours.fr}

\affiliation[a]{
Institut Denis Poisson, UMR - CNRS 7013, \\
Universit\'{e} de Tours, Parc de Grandmont, 37200 Tours, France
}

\affiliation[b]{
Department of General Relativity and Gravitation, Institute of Physics,\\
Kazan Federal University, Kremlevskaya street 18, 420008 Kazan, Russia
}


\abstract{
We present the consistent theory of a free massive spin-2 field with 5  degrees of freedom propagating
in spacetimes  with an arbitrary geometry. We obtain this theory  via 
linearizing the equations of the ghost-free massive gravity expressed in the tetrad formalism. 
The theory is parameterized by a {\it non-symmetric} rank-2 tensor whose 16 components fulfill 11 constraints 
implied by the equations. 
When restricted to Einstein spaces, 
the theory reproduces the standard description of massive gravitons. 
 In generic spacetimes, the theory  does not show the massless limit and 
 always propagates five degrees of freedom, even for the vanishing mass parameter. 
 We illustrate these features by an explicit calculation for a homogeneous and isotropic 
 cosmological background. It turns out  that the spin-2 particles are always stable if they are sufficiently massive, 
 hence they may be a part of the Dark Mater. 
}


\maketitle

\noindent 
\section{Introduction}
Equations of massive fields of spin $0,1/2,1,3/2$ in Minkowski space (the Klein-Gordon, Dirac, Proca,
Rarita-Schwinger)  directly generalize to curved space, but for the massive spin-2
field this does not work.  The Fierz-Pauli (FP) theory of massive gravitons 
 \cite{Fierz:1939ix}
generalises to curved space only for special  spacetimes: Einstein spaces, whose Ricci tensor 
is proportional to the metric, $R_{\mu\nu}=\Lambda g_{\mu\nu}$ 
 \cite{Aragone:1971kh,Aragone:1979bm,Higuchi}. In an arbitrary spacetime  the theory 
 shows six instead of five dynamical graviton polarizations, the extra polarization state being ghost-type.
 This feature 
 was for a longtime  thought  to be inevitable \cite{Buchbinder:1999ar}, hence 
 all applications of the massive spin-2 field have been limited 
 only to Einstein spaces. 
 
 Quite recently,  a consistent theory of a free massive spin-2 field has  nevertheless  been constructed
 \cite{Bernard:2014bfa,Bernard:2015mkk,Bernard:2015uic} 
 by applying the methods of 
 the dRGT ghost-free massive gravity theory \cite{deRham:2010kj}\footnote{dRGT -- after the names of the authors of \cite{deRham:2010kj}.}.
 The dRGT  theory propagates  5 polarizations  at the non-linear level and 
 contains  the physical metric $g_{\mu\nu}$ and a reference metric $f_{\mu\nu}$. 
 The procedure of \cite{Bernard:2014bfa,Bernard:2015mkk,Bernard:2015uic}
 was to linearize the dRGT field equations with respect to the perturbations 
 $\delta g_{\mu\nu}\equiv h_{\mu\nu}$ and then replace in the obtained linear equations 
 the reference metric  $f_{\mu\nu}$ by the
  expression obtained by resolving the background non-linear equations with respect to $f_{\mu\nu}$. This yields linear equations for the symmetric tensor $h_{\mu\nu}$ 
 and the coefficients in these equations are determined only by the background metric $g_{\mu\nu}$.  
 For any background these equations imply 5 constraints reducing the number of propagating 
 degrees of freedom (DoF) to $5=10-5$ \cite{Bernard:2014bfa,Bernard:2015mkk,Bernard:2015uic}. 
 
 Unfortunately, the mass term for $h_{\mu\nu}$ obtained in 
 \cite{Bernard:2014bfa,Bernard:2015mkk,Bernard:2015uic}  is rather complicated 
 and even the very demonstration of the existence of the scalar constraint removing the sixth polarization requires 
 tedious calculations. This can be traced to the following fact. The dRGT mass term \cite{deRham:2010kj} 
 is expressed in terms of the tensor $\gamma^\mu_{~\nu}$ defined by the conditions 
 \be
 \gamma^\mu_{~\sigma}\gamma^\sigma_{~\nu}=g^{\mu\sigma}f_{\sigma\nu}\, .
 \ee
Linearizing  this with respect to $\delta g_{\mu\nu}$ yields 
\be                \label{dg}
 \delta\gamma^\mu_{~\sigma}\gamma^\sigma_{~\nu}
 +\gamma^\mu_{~\sigma}\delta \gamma^\sigma_{~\nu}
 =\delta g^{\mu\sigma}f_{\sigma\nu}\,,
 \ee
 which can be viewed as the so-called Sylvestre matrix equation determining 
 $\delta\gamma^\mu_{~\sigma}$ in terms of $\delta g_{\mu\nu}$. Its solution 
 exists (generically) but is rather involved, 
 which is why the mass term in the resulting equations has a very complicated structure
  \cite{Bernard:2014bfa,Bernard:2015mkk,Bernard:2015uic}. 
 
 One may think that the situation can be improved by choosing  $\delta\gamma^\mu_{~\sigma}$ 
 as  fundamental variables describing  the perturbations and then to use \eqref{dg} to express 
 $\delta g_{\mu\nu}$ in terms of $\delta\gamma^\mu_{~\sigma}$. However, the {\it kinetic} 
 term of the resulting equations turns out to be very complicated  in this case.   Therefore,
 a different approach is needed. 
 
 In what follows,  we reconsider the procedure of
 \cite{Bernard:2014bfa,Bernard:2015mkk,Bernard:2015uic} 
 within the {\it tetrad formulation} of the dRGT theory\footnote{The linearization of the tetrad version of the dRGT theory 
 within the Palatini approach was considered  in  \cite{Deser:2015wta}.}
 \cite{Chamseddine:2011mu,Hinterbichler:2012cn}
 instead of the metric formulation considered in \cite{Bernard:2014bfa,Bernard:2015mkk,Bernard:2015uic}. 
 Within the {tetrad formulation},  
 the two metrics of the dRGT theory are decomposed with respect to tetrads $e^a_{~\mu}$ and $\f^a_{~\mu}$ 
 (see Eq.\eqref{tet} below) and one has 
 \be                  \label{d2}
 \gamma^\mu_{~\nu}=e_a^{~\mu}\f^a_{~\nu}.
 \ee
 The idea is to linearize the dRGT equations with respect to {perturbations of the physical tetrad} 
 $\delta e^a_{~\mu}$.  
 Eq.\eqref{d2} then yields a very simple expression for $\delta  \gamma^\mu_{~\nu}$ in terms of $\delta e^a_{~\mu}$,
 which leads in the end to a simple form of the resulting linear equations for $\delta e^a_{~\mu}$. 
 These equations can be  reformulated entirely  in terms of 
  the {\it non-symmetric} tensor 
 \be
 X_{\mu\nu}=\eta_{ab}\, e^a_{~\mu}\delta e^b_{~\nu}
 \ee
which  is related to the symmetric tensor used in \cite{Bernard:2014bfa,Bernard:2015mkk,Bernard:2015uic} 
via $h_{\mu\nu}=X_{\mu\nu}+X_{\nu\mu}$. 

After having derived  the equations for $X_{\mu\nu}$, 
we can completely forget their  tetrad origin and use only the standard 
tensor language. 
As a result, we obtain linear equations for a non-symmetric tensor field 
$X_{\mu\nu}$ propagating  in a spacetime with the metric  $g_{\mu\nu}$. 
We use these equations to 
describe the massive spin-2 field.  At first glance, 
using non-symmetric tensors may seem odd.  However, our equations  turn out to be quite simple 
and they immediately imply 11 contraints, hence  among 16 components $X_{\mu\nu}$ 
there are only $5=16-11$ independent ones. This matches the number of polarizations of massive spin-2 particles. 
In particular, the absence of the sixth 
 polarization is easy to see in our theory.

 The consistency of the linearization  procedure  requires that the background dRGT equations should be fulfilled
 as well, which   can be achieved by adjusting the reference metric, 
 hence $f_{\mu\nu}$ becomes a function of $g_{\mu\nu}$. 
 This gives a consistent theory of  the massive spin-2 field  for any $g_{\mu\nu}$. 
 The theory is not unique since the background dRGT equations, viewed as algebraic equations 
 for $f_{\mu\nu}$, may have several  solutions 
 for a given $g_{\mu\nu}$. This determines several possible mass terms, hence several 
 different theories. 
 In general, the mass term  depends non-linearly on the background Ricci tensor $R_{\mu\nu}$, but 
 there exist special cases in which this dependence is linear. 
 This corresponds to two special theories, 
  called below models I and II. 
  
  Summarizing, we shall present in what follows a consistent theory for  a free massive spin-2 field in an 
  arbitrary spacetime expressed in terms of a non-symmetric tensor $X_{\mu\nu}$.
  Our theory turns out to be simpler than the one of 
  \cite{Bernard:2014bfa,Bernard:2015mkk,Bernard:2015uic}  
 expressed   in terms of $h_{\mu\nu}=X_{\mu\nu}+X_{\nu\mu}$. 
  Our theory and the one of \cite{Bernard:2014bfa,Bernard:2015mkk,Bernard:2015uic} 
    are presumably equivalent
  since they are both obtained in a similar way from the dRGT theory, 
  but the equivalence is not manifest   since 
  it is difficult to perform the  inverse transformation to express $X_{\mu\nu}$ in terms of $h_{\mu\nu}$. 
It would  probably be correct to say that the two theories should be equivalent {\it generically}, up to 
exceptional cases where the inverse transformation from $h_{\mu\nu}$ to $X_{\mu\nu}$ degenerates. 
This corresponds to the fact that the metric and tetrad formulations of the dRGT theory are 
equivalent  generically, but the equivalence may be lost for exceptional parameter values comprising a 
zero measure set in the whole parameter  space \cite{Deffayet:2012zc}. 

Although one may view our  spin-2 particles  as massive gravitons,  we rather prefer associate gravitons with the quanta 
of the background metric  $g_{\mu\nu}$.  In fact, after having derived its equations, we may 
totally forget  
about its relation  to gravity 
and  consider the $X_{\mu\nu}$ field 
as describing  spin-2 massive  ``mesons" propagating in a curved  spacetime. Depending on our choice, this field 
may or may not interact with other matter fields, although its always interacts with the 
background gravity. 
Our primary goal was to  construct the consistent 
mathematical description for this field, while its  interpretation and possible physical applications may be different.

 The rest of this text is organized as follow. In Sections II and III  we present  the equations 
 of the dRGT massive gravity in the tetrad formalism and describe their  linearization with respect to the tetrad 
 perturbations.  The tetrads appear in the  coefficients of the resulting linear equations,
 but they can be eliminated  by using the background field equations. 
 As a result, after having used  them as the technical tool, we can get rid of the tetrads altogether
 and consider the  theory of a non-symmetric tensor field $X_{\mu\nu}$ 
propagating   in a curved  spacetime. 
 Section IV  contains  the analysis of constraints  implied by 
 the equations for  $X_{\mu\nu}$ 
  and shows that there are only 5 propagating DoF. 
  Section V presents  two simple
 versions of the theory,  called model I and model II,  for which the mass term is  linear in $R_{\mu\nu}$. 
 Sections VI and VII  show how  these two models behave in Einstein spaces and in the spatially flat 
 Friedmann-Lema$\hat{{\i}}$tre-Robertson-Walker (FLRW) universe. 
 In the latter case the massive spin-2 particles  are found to be stable, at least at late times, hence they 
 could potentially contribute to the Dark Matter. The  backreaction of the massive spin-2 field on the 
 background geometry 
 is discussed in Section VIII, while Section IX contains a brief summary of results. 
 Many technical details are explained in the 
 five Appendices.  
 
 A short version of this text  can be found in  \cite{Mazuet:2017hey}.

\section{The dRGT massive gravity}
\setcounter{equation}{0}

The dRGT theory \cite{deRham:2010kj} is expressed in terms of  the dynamical  spacetime metric $g_{\mu\nu}$
and a non-dynamical  reference metric $f_{\mu\nu}$. The latter can be arbitrary,  for example 
 flat, but it is convenient for our purposes not to specify it for the time being.  

The two metrics can be decomposed with respect 
to two  tetrads $e^a_{~\mu}$ and $\f^a_{~\mu}$  as 
\be                        \label{tet}
g_{\mu\nu}=\eta_{ab}\,e^a_{~\mu}e^b_{~\nu},~~~~~~~~~~
f_{\mu\nu}=\eta_{ab}\,\f^a_{~\mu}\f^b_{~\nu}, 
\ee 
where 
$\eta_{ab}={\rm diag}[-1,1,1,1]$. The inverse metrics are
\be
g^{\mu\nu}=\eta^{ab}\,e_a^{~\mu}e_b^{~\nu},~~~~~~~~~
f^{\mu\nu}=\eta^{ab}\,\f_a^{~\mu}\f_b^{~\nu}\,,
\ee 
where 
\be
e^a_{~\mu} e_b^{~\mu}=\delta^a_b,~~~~
e^a_{~\mu} e_a^{~\nu}=\delta^\nu_\mu,~~~~~~~
\f^a_{~\mu} \f_b^{~\mu}=\delta^a_b,~~~~
\f^a_{~\mu} \f_a^{~\nu}=\delta^\nu_\mu.
\ee 
  One denotes 
\be
|e^a_{~\mu}|\equiv e=\sqrt{-g},~~~~~~~~~~~
|\f^a_{~\mu}|\equiv \f=\sqrt{-f}. 
\ee
The action of the theory is 
\be
S_{\rm dRGT}=M_{\rm Pl}^2 \int \left(\frac12\,R(g)\sqrt{-g}
- U  \right)d^4 x\equiv M_{\rm Pl}^2\int L_{\rm dRGT}\, d^4x\,,
\ee
where $M_{\rm Pl}$ is the Planck mass and 
the potential  is $U=U_0+U_1+U_2+U_3$
with
\be
{U}_0&=&\beta_0\,\frac{1}{4!}\,
\epsilon_{abcd}\,\epsilon^{\mu\nu\alpha\beta}e^a_{~\mu}e^b_{~\nu}
e^c_{~\alpha}e^d_{~\beta} \,,\nonumber \\
{U}_1&=&\beta_1\,\frac{1}{3!}\,
\epsilon_{abcd}\,\epsilon^{\mu\nu\alpha\beta}e^a_{~\mu}e^b_{~\nu}
e^c_{~\alpha}\f^d_{~\beta}   \,,    \nonumber \\
{U}_2&=&\beta_2\,\frac{1}{2!2!}\,
\epsilon_{abcd}\,\epsilon^{\mu\nu\alpha\beta}e^a_{~\mu}e^b_{~\nu}
\f^c_{~\alpha}\f^d_{~\beta}  \,,\nonumber \\
{U}_3&=&\beta_3\,\frac{1}{3!}\,
\epsilon_{abcd}\,\epsilon^{\mu\nu\alpha\beta}e^a_{~\mu}\f^b_{~\nu}
\f^c_{~\alpha}\f^d_{~\beta}\,.
\ee
Here $\beta_A$ are parameters with the dimension $(\mbox{mass})^2$;  we  assume 
$\epsilon_{0123}=\epsilon^{0123}=+1$. 

Let us vary the action with respect to the tetrad $e^a_{~\mu}$. One has 
\be
\delta g_{\mu\nu}\equiv h_{\mu\nu}=
\delta e^a_{~\mu} \,e_{a\nu}+e_{a\mu}\, \delta e^a_{~\nu}~~~~
\ee
hence 
\be
\delta\left(\frac12\, R(g)\sqrt{-g}\right)
=\frac12 \, G_{\mu\nu}\delta g^{\mu\nu} \sqrt{-g}
=-\frac12 \, G^{\mu\nu}\delta g_{\mu\nu} \sqrt{-g}
=-e\, G_a^{~\mu}\delta\, e^a_{~\mu}
\ee
where $G_a^{~\mu}=e_a^{~\sigma} G_\sigma^{~\mu}$. 
To vary the potential $U$ one uses the relations 
\be
\frac{1}{4!}\,
\epsilon_{abcd}\epsilon^{\mu\nu\alpha\beta}e^a_{~\mu}e^b_{~\nu}
e^c_{~\alpha}e^d_{~\beta}&=&e\,,\nonumber \\
\frac{1}{3!}\,
\epsilon_{abcd}\epsilon^{\mu\nu\alpha\beta}e^b_{~\nu}
e^c_{~\alpha}e^d_{~\beta}&=&e\, e_a^{~\mu}\,, \nonumber \\
\frac{1}{2!}\,
\epsilon_{abcd}\epsilon^{\mu\nu\alpha\beta}
e^c_{~\alpha}e^d_{~\beta}&=&e\, (e_a^{~\mu}e_b^{~\nu} 
-e_a^{~\nu}e_b^{~\mu})\,. 
\nonumber 
\ee
This yields, for example, 
\be
\delta {U}_0&=&\beta_0\,\frac{1}{3!}\,
\epsilon_{abcd}\epsilon^{\mu\nu\alpha\beta}\delta e^a_{~\mu}\, e^b_{~\nu}
e^c_{~\alpha}e^d_{~\beta} =\beta_0 \, e\, e_a^{~\mu}\delta e^a_{~\mu}
\ee
and similarly for the other $\delta U_A$. As a result, one obtains 
\be
\delta L_{\rm dRGT}=-e\,{\bf E}_a^{~\mu} \delta e^a_{~\mu}=-e\,{\bf E}_{ab}\,e^b_{~\mu} \delta e^a_{~\mu}
\ee
where
\be                    \label{E1}
{\bf E}_{ab}&\equiv & G_{ab}+\M_{ab}
\ee
with  $G_{ab}=e_a^{~\mu} e_b^{~\nu} G_{\mu\nu}$ and 
\be                  \label{EE1}
\M_{ab}=\M^{(0)}_{ab}+\M^{(1)}_{ab}
+\M^{(2)}_{ab}
+\M^{(3)}_{ab}
\ee
with 
\be                   \label{MM}
\M^{(0)}_{ab}&=&\beta_0\,\eta_{ab}, \nn \\
\M^{(1)}_{ab}&=&\beta_1([\gm]\,\eta_{ab}-\gm_{ab}), \nn \\
\M^{(2)}_{ab}&=&\beta_2\,|\gamma|\,\left( [\gam]\,\gam_{ab}  - (\gam^2)_{ab}  \right), \nn \\
\M^{(3)}_{ab}&=&\beta_3\,|\gamma|\,\gam_{ab}.
\ee
Here
we introduced the mutually inverse matrices 
\be
\gm^a_{~b}=\f^a_{~\sigma} e_b^{~\sigma}\,,~~~~~~~
\gam^a_{~b}=e^a_{~\sigma} \f_b^{~\sigma}\,,~~~~~~~~\gm^a_{~b}\gam^b_{~c}=\delta^a_c\,,
\ee
and denoted the determinant and trace\footnote{Later we shall sometimes use the hat 
for matrices, for example $\hat{\gamma}=\gamma^a_{~b}$, but we shall not always write the hat 
under the trace sign,  hence  $[\hat{\gamma}]\equiv [\gm]=\gm^a_{~a}$.}
as 
\be
|\gamma|\equiv \det(\gamma^a_{~b})=\frac{\phi}{e},~~~~~~~~~[\gm]\equiv \gm^a_{~a}\,.
\ee
The matrix indices are moved by $\eta_{ab}$, for example  $\gam_{ab}=\eta_{ac}\gam^c_{~b}$ and 
$(\gam^2)_{ab}=\gam_{ac}\gam^c_{~b}$.  

Vanishing of the  variation of the action gives the dRGT 
field equations, 
\be                  \label{E2}
{\bf E}_{ab}\equiv G_{ab}+{\rm M}_{ab}=0,
\ee
or explicitly 
\be                       \label{dRGT}
G_{ab}+\beta_0\,\eta_{ab}+\beta_1([\gm]\,\eta_{ab}-\gm_{ab})+
\beta_2\,|\gamma|\left( [\gam]\,\gam_{ab}  - (\gam^2)_{ab}  \right)+\beta_3\,|\gamma|\,\gam_{ab}=0.
\ee
Taking their antisymmetric part yields 
\be                                   \label{symm} 
\beta_1\,\gm_{[ab]}=(\beta_2+\beta_3)|\gamma| \gam_{[ab]}-\beta_2|\gamma| (\gam^2)_{[ab]}.
\ee
Since the matrices  $\gm_{ab}$ and $\gam_{ab}$ are position-dependent, these conditions can be 
fulfilled in the generic case if only 
$\gm_{[ab]}=0$ and $\gam_{[ab]}=0$. Therefore, matrices $\gm_{ab}$ and $\gam_{ab}$ are symmetric,
\be              \label{sym0}
\gm_{ab}=\gm_{ba},~~~~~~~\gam_{ab}=\gam_{ba}. 
\ee
Exceptionally, for special values of the parameters $\beta_A$, 
there could be special solutions of the equations for which $\gm_{[ab]}\neq 0$ and $\gam_{[ab]}\neq 0$
but the conditions \eqref{symm} are still fulfilled\footnote{If $\gm_{[ab]}\neq 0$ then the metric and tetrad formulations 
 of the dRGT theory are not equivalent.}. 
 However, such cases are very special 
 \cite{Deffayet:2012zc}
while for generic solutions of the equations the matrices  $\gm_{ab}$ and $\gam_{ab}$ are symmetric,
which we shall always assume in what follows.

 \section{Equations for perturbations}
 \setcounter{equation}{0}
 
 Let $e^a_{~\mu}$ be a solution of ${\bf E}_{ab}=0$ and 
 consider a perturbed configuration
 $e^a_{~\mu}\to e^a_{~\mu}+\delta e^a_{~\mu}$. Assuming that the latter 
 also fulfills the equations and expanding ${\bf E}_{ab}$ in terms of $\delta e^a_{~\mu}$ yields 
 \be         \label{exp}
0={\bf E}_{ab} =\overset{(0)}{{\bf E}}_{ab}+\overset{(1)}{{\bf E}}_{ab}+\ldots
 \ee
 where $\overset{(0)}{{\bf E}}_{ab}$, 
 $\overset{(1)}{{\bf E}}_{ab}$
  and the dots denote, respectively, terms which are zeroth order, first order, 
 and higher order in $\delta e^a_{~\mu}$. 
 The zeroth order term vanishes
 since, by assumption, $e^a_{~\mu}$ fulfills the 
 equations,  hence $\overset{(0)}{{\bf E}}_{ab}=0$. 
 Therefore, neglecting the higher order terms, the equations reduce to 
 \be
\overset{(1)}{{\bf E}}_{ab}\equiv \delta {\bf E}_{ab}=0. 
 \ee
 To compute $\delta {\bf E}_{ab}$ one 
represents the tetrad perturbation  as 
\be
\delta e^a_{~\mu}=X^a_{~b} \,e^b_{~\mu}. 
\ee
It follows that 
\be
X^a_{~b} =e_b^{~\mu}\delta e^a_{~\mu} 
\ee
hence the 16 coefficients $X^a_{~b}$ are the tetrad perturbations 
projected on the unperturbed tetrad. 
The symmetric part of $X_{ab}=\eta_{ac}X^c_{~b}$ 
determines perturbations of the metric, 
\be
e_a^{~\mu}e_b^{~\mu}\,\delta g_{\mu\nu}\equiv h_{ab}=X_{ab}+X_{ba}.
\ee
It follows also that 
\be                           \label{gg}
\delta e_a^{~\mu}&=&-X^m_{~~a}\,e_m^{~\mu}\,, ~~~~~~~~~~
\delta |\gamma| =-[X]\,|\gamma|\,,   \nn \\
\delta\gm^a_{~b}&=&-\gm^a_{~m} X^m_{~~b}\,,  ~~~~~~~~~
\delta \gam^a_{~b}=X^a_{~m}\gam^m_{~~b}\,,
\ee
where $[X]=X^a_{~a}$. As a result, 
the perturbation equations read 
\be                   \label{M0}
\delta {\bf E}_{ab}\equiv \delta G_{ab}+\delta \M_{ab}=0, 
\ee
where $\delta G_{ab}$ are  perturbations of the tetrad projections of the Einstein tensor and 
\be                   \label{M00}
\delta \M_{ab}=\delta \M^{(1)}_{ab}+\delta \M^{(2)}_{ab}+\delta \M^{(3)}_{ab}\,,
\ee 
where $\delta \M^{(A)}_{ab}$ are obtained by perturbing  the  $\M^{(A)}_{ab}$ in \eqref{MM}:
\be                 \label{M}
\delta \M^{(1)}_{ab}&=&\beta_1\left(
\gm_{am}X^m_{~~b}-\eta_{ab}\,\gm^{mn}\,X_{mn}\right)  \,, \nn \\
\delta \M^{(2)}_{ab}&=&\beta_2\,|\gamma|\,\{((\gam^2)_{ab}-[\gam]\gam_{ab})[X]
+\gam_{ab}\,\gam^{mn}\,X_{mn}   \nn \\
&&+[\gam]\,X_{an}\,\gam^n_{~~b}
-X_{an}(\gam^2)^n_{~~b}-\gam_{am}\,X^m_{~~n}\,\gam^n_{~~b} \}\,, \nn \\
\delta \M^{(3)}_{ab}&=&\beta_3\,|\gamma|\,\left(
X_{am}\gam^m_{~~b}-[X]\gam_{ab}
\right),
\ee
whereas $\delta \M^{(0)}_{ab}=0$. 

We shall later need $\M^{(2)}_{ab}$ and 
$\delta \M^{(2)}_{ab}$  and also  $\M^{(3)}_{ab}$ and 
$\delta \M^{(3)}_{ab}$
expressed entirely in terms of $\gm_{ab}$ instead of $\gam_{ab}$. 
Such expressions can be obtained by 
applying the 
 Hamilton-Cayley  relation valid for any $4\times 4$ matrix $\hat{A}$: 
\be                           \label{A}
\e_0(\hat{A})\,\hat{A^4}-\e_1(\hat{A})\,\hat{A}^3+\e_2(\hat{A})\,\hat{A}^2-\e_3(\hat{A})\,\hat{A}+\e_4(\hat{A})=0,
\ee
where $\e_0(\hat{A})=1$ while the other coefficients are the symmetric 
polynomials of the eigenvalues $\lambda_a$ of $\hat{A}$, 
\be                        \label{inv}
\e_1(\hat{A})&=&[\hat{A}]=\sum_a\lambda_a,
~~~~~~~~~~~~\e_2(\hat{A})=\frac12([\hat{A}]^2-[\hat{A}^2])=\sum_{a<b}\lambda_b\lambda_b ,  \\
\e_3(\hat{A})&=&\frac16([\hat{A}]^3-3[\hat{A}][\hat{A}^2]
+2[\hat{A}^3])=\sum_{a<b<c}\lambda_b\lambda_b\lambda_c,~~~~
\e_4(\hat{A})=\det(\hat{A})=\lambda_1\lambda_2\lambda_3\lambda_4.  \nn
\ee           \label{Meq0}
One has 
\be
\frac{\e_k(\hat{A})}{\e_4(\hat{A})}=\e_{4-k}(\hat{A}^{-1}),~~~~~~~k=0,1,2,3,4. 
\ee
Multiplying Eq.\eqref{A} by  $\hat{A}^{-2}/\e_4(\hat{A})$ yields 
\be
\frac{1}{\e_4(\hat{A})}([\hat{A}]\hat{A}-\hat{A}^2)=\frac{\e_2(\hat{A})}{\e_4(\hat{A})}
-\frac{\e_3(\hat{A})}{\e_4(\hat{A})}\,\hat{A}^{-1}+\hat{A}^{-2}
\ee
and hence 
\be
{\e_4(\hat{A}^{-1})}([\hat{A}]\hat{A}-\hat{A}^2)=\e_2(\hat{A}^{-1})-\e_1(\hat{A}^{-1})\,\hat{A}^{-1}+\hat{A}^{-2}. 
\ee
Applying this to $\hat{A}=\gam^a_{~b}$ and  $\hat{A}^{-1}=\gm^a_{~b}$ allows one to express 
$\M^{(2)}_{ab}$ in \eqref{MM} as 
\be    \label{id}
\frac{1}{\beta_2}\,\M^{(2)}_{ab}&=&
|\gamma |([\gam]\gam_{ab}-(\gam^2)_{ab}) \nn   \\
&=&(\gm^2)_{ab}-[\gm]\gm_{ab}+\frac12([\gm]^2-[\gm^2])\,\eta_{ab}\,.
\ee
Varying the expression in the second line here yields 
\be             \label{Meq}
\frac{1}{\beta_2}\,\delta \M^{(2)}_{ab}=&-&\gm^m_{~a}\gm^n_{~b}\,X_{mn}-(\gm^2)^m_{~a}\,X_{mb}  \nn \\
&+&\gm_{ab}\,\gm_{mn}\,X^{mn}+[\gm]\,\gm^m_{~a}\,X_{mb} \nn \\
&+&((\gm^2)_{mn}\,X^{mn}-[\gm]\,\gm_{mn}\,X^{mn})\,\eta_{ab}.
\ee
Similar manipulations with the Hamilton-Cayley relation yield 
\be    \label{id1}
\frac{1}{\beta_3}\,\M^{(3)}_{ab}&=& |\gamma|\gam_{ab}=-(\gm^3)_{ab}
+\e_1(\hat{\gm})\,(\gm^2)_{ab}-\e_2(\hat{\gm})\,\gm_{ab}+\e_3(\hat{\gm})\,\eta_{ab}\,,
\ee
which determines also the coefficients in $\delta\M^{(3)}_{ab}$  in \eqref{M}. 

\subsection{Eliminating the tetrads}
Summurizing the above discussion, the equations for the tetrad perturbations are given 
by \eqref{M0}--\eqref{M}. They have been obtained within the tetrad formalism 
and they are expressed in terms of tetrad projections. 
However, after having obtained these equations, 
we can now eliminate the tetrads altogether from their coefficients and pass to the 
standard tensorial description. 
The first step is to transform the equations to 
\be
E_{\mu\nu}\equiv e^a_{~\mu}e^b_{~\mu}(\delta G_{ab}+\delta \M_{ab})\equiv \Delta_{\mu\nu}+{\cal M}_{\mu\nu}=0.
\ee
The kinetic term here is 
\be            \label{del}
\Delta_{\mu\nu}\equiv  e^a_{~\mu}e^b_{~\mu}\,\delta G_{ab}&=&
 e^a_{~\mu}e^b_{~\mu}\,\delta(G_{\rho\sigma}e_a^{~\rho}e_{b}^{~\sigma}) \nn \\
&=&e^a_{~\mu}e^b_{~\mu}\,\left(e_a^{~\rho}e_{b}^{~\sigma} \,\delta G_{\rho\sigma}
+G_{\rho\sigma}e_b^{~\sigma}\delta e_{a}^{~\rho}+G_{\rho\sigma}e_a^{~\rho}\delta e_{b}^{~\sigma}\right) \nn \\
&=&e^a_{~\mu}e^b_{~\mu}\,\left(e_a^{~\rho}e_{b}^{~\sigma} \,\delta G_{\rho\sigma}
-G_{mb}X^m_{~~a}-G_{am}X^m_{~~b}\right)  \nn   \\
&=& \delta G_{\mu\nu}-G_{\mu\sigma}X^\sigma_{~\nu}-G_{\nu\sigma}X^\sigma_{~\mu}\,,
\ee
where
\be
X^\mu_{~\nu}
\equiv X^a_{~b}\,e_a^{~\mu}e^b_{~\nu}=e_a^{~\mu}\delta e^a_{~\nu}\,.
\ee
The variation of the Einstein tensor $\delta G_{\mu\nu}$ 
in terms of $h_{\mu\nu}=\delta g_{\mu\nu}$ is well known, 
\be
2\,\delta G_{\mu\nu}=
\nabla^\sigma\nabla_\mu \h_{\nu\sigma}+\nabla^\sigma\nabla_\nu \h_{\mu\sigma}
-\Box \h_{\mu\nu}-\nabla_{\mu}\nabla_\nu\h^\alpha_{~\alpha}  \nn \\
+g_{\mu\nu}(\Box \h^\alpha_{~\alpha}-\nabla^\alpha\nabla^\beta h_{\alpha\beta}+R^{\alpha\beta} \h_{\alpha\beta})
-R\,\h_{\mu\nu}, 
\ee
where   $\nabla_\mu$ is the usual covariant derivative with respect to the 
Christoffel connection.  At the same time, one has 
\be
h_{\mu\nu}=e^a_{~\mu}e^b_{~\nu}(X_{ab}+X_{ba})=X_{\mu\nu}+X_{\nu\mu}
\ee
where 
\be                        \label{X}
X_{\mu\nu}=g_{\mu\sigma}X^\sigma_{~\nu}=\eta_{ab}\, e^a_\mu\,\delta e^b_\nu\,.
\ee
Inserting everything to \eqref{del} yields the 
kinetic operator in the form not containing any reference to the tetrads, 
\be                    \label{Del}
\Delta_{\mu\nu}&=&\frac12\nabla^\sigma\nabla_\mu (X_{\sigma\nu} +X_{\nu\sigma})
 +\frac12\nabla^\sigma\nabla_\nu (X_{\sigma\mu} +X_{\mu\sigma}) \nn \\
&-&\frac12\Box ({X}_{\mu\nu}+X_{\nu\mu})-\nabla_\mu\nabla_\nu{[X]}  \nn \\
 &+&g_{\mu\nu}\left(\Box [X]-\nabla^\alpha\nabla^\beta X_{\alpha\beta}+R^{\alpha\beta}X_{\alpha\beta}
\right) \nn  \\
& -&R^\sigma_\mu X_{\sigma\nu}-R^\sigma_\nu X_{\sigma\mu}\,,
\ee
with $[X]=X^\alpha_{~\alpha}$. 

Next,  the mass term is 
\be                             \label{Mb0}
{\cal M}_{\mu\nu}&=&e^a_{~\mu}e^b_{~\nu}\delta \M_{ab}=
e^a_{~\mu}e^b_{~\nu}\,( \delta \M^{(1)}_{ab}+\delta \M^{(2)}_{ab}+\delta \M^{(3)}_{ab})  \nn \\
&\equiv & {\cal M}^{(1)}_{\mu\nu}+ {\cal M}^{(2)}_{\mu\nu}+{\cal M}^{(3)}_{\mu\nu}\
\ee
where, using \eqref{M} and \eqref{Meq}, 
\be                 \label{Mb}
{\cal M}^{(1)}_{\mu\nu}&=&\beta_1\left(
\gm_{~\mu}^\sigma X_{\sigma\nu}-g_{\mu\nu}\,\gm^{\alpha\beta}\,X_{\alpha\beta}\right)  \,, \nn \\
{\cal M}^{(2)}_{\mu\nu}&=&\beta_2\,\{ -\gm^\alpha_{~\mu}\gm^\beta_{~\nu}\,X_{\alpha\beta}
-(\gm^2)^\alpha_{~\mu}\,X_{\alpha\nu} 
+\gm_{\mu\nu}\,\gm_{\alpha\beta}\,X^{\alpha\beta} \nn \\
&&+[\gm]\,\gm^\alpha_{~\beta}\,X_{\alpha\nu} +((\gm^2)_{\alpha\beta}\,X^{\alpha\beta}
-[\gm]\,\gm_{\alpha\beta}\,X^{\alpha\beta})\, g_{\mu\nu}\}
\,, \nn \\
{\cal M}^{(3)}_{\mu\nu}&=&\beta_3\,|\gamma|\,\left(
X_{\mu\sigma}\gam^\sigma_{~\nu}-[X]\gam_{\mu\nu}
\right). 
\ee
Using Eq.\eqref{XXX} below, ${\cal M}^{(3)}_{\mu\nu}$ can be expressed entirely in terms of $\gamma_{\mu\nu}$,
but we shall rather need it expressed in terms of $\gam_{\mu\nu}$.

The coefficients in \eqref{Mb}  still depend on the tetrads via  combinations
\be
\gm^\mu_{~\nu}=e_a^{~\mu}\f^a_{~\nu}\,,~~~|\gamma|=\det(\gm^\mu_{~\nu} )=e_4(\gamma^\mu_{~\nu}),~~~\nn \\
\gm_{\mu\nu}=g_{\mu\sigma}\gm^\sigma_{~\nu},  ~~~
\gam^\mu_{~\nu}=\f_a^{~\mu} e^a_{~\nu}\, , 
~~~
\gam_{\mu\nu}=g_{\mu\sigma}\gam^\sigma_{~\nu}.  
\ee
Now, the crucial point is that these quantities can be obtained from the background equations. 
Let us remember that we are expanding the field equations as expressed by \eqref{exp}
and that the zeroth order term in this expansion should vanish for the procedure 
to be consistent. Hence 
the background equations should be fulfilled.  We also remember that 
up to now  $\gm_{\mu\nu}$ and $\gam_{\mu\nu}$  have essentially remained undetermined  since 
the tetrad $\f_a^{~\mu}$ has never been specified. On the other hand, 
the background dRGT equations \eqref{dRGT} read
\be                         \label{EG} 
{\bf E}_{\mu\nu}\equiv G_{\mu\nu}+\beta_0\,g_{\mu\nu}+\beta_1([\gm]\,g_{\mu\nu}-\gm_{\mu\nu})
+\beta_2\,|\gamma|\left( [\gam]\,\gam_{\mu\nu}  - (\gam^2)_{\mu\nu}  \right)
+\beta_3\,|\gamma|\,\gam_{\mu\nu}=0, ~~~~~
\ee
and these can be viewed as {\it algebraic  conditions} determining  $\gm_{\mu\nu}$ and $\gam_{\mu\nu}$   in terms of the 
background metric $g_{\mu\nu}$ and its Einstein tensor $G_{\mu\nu}$. The idea therefore is to fulfill the background 
equations not by solving them for $g_{\mu\nu}$ but by adjusting $\gm_{\mu\nu}$, $\gam_{\mu\nu}$  
for a given $g_{\mu\nu}$. 

These equations can also be represented as follows. 
The identities  \eqref{id} and \eqref{id1} 
yield 
\be                  \label{XXX}
|\gamma|\left( [\gam]\,\gam_{\mu\nu}  - (\gam^2)_{\mu\nu}  \right)&=&(\gm)^2_{\mu\nu}
-e_1(\hat{\gm})\,\gm_{\mu\nu}+e_2(\hat{\gm})\,g_{\mu\nu}\,,   \nn \\
|\gamma|\gam_{\mu\nu}&=&-(\gm^3)_{\mu\nu}
+e_1(\hat{\gm})\,(\gm^2)_{\mu\nu}-e_2(\hat{\gm})\,\gm_{\mu\nu}+e_3(\hat{\gm})\,g_{\mu\nu}\,,
\ee
where $e_A(\hat{\gm})\equiv e_A(\gm^\mu_{~\nu})$. In view of this, 
\eqref{EG} can be represented in the form containing only $\gm_{\mu\nu}$, 
\be                         \label{EG1} 
{\bf E}_{\mu\nu}\equiv G_{\mu\nu}+\beta_0\,g_{\mu\nu}&+&\beta_1(e_1(\hat{\gm})\,g_{\mu\nu}-\gm_{\mu\nu})
+\beta_2\,\left((\gm^2)_{\mu\nu}-e_1(\hat{\gm})\gm_{\mu\nu}+ e_2(\hat{\gm})g_{\mu\nu}\right)  \nn \\
&+&\beta_3\left(-(\gm^3)_{\mu\nu}+e_1(\hat{\gm})(\gm^2)_{\mu\nu}
-e_2(\hat{\gm})\gm_{\mu\nu}+e_3(\hat{\gm})g_{\mu\nu}\right)=0. ~~~~~
\ee
For any value of the background metric $g_{\mu\nu}$, these can be viewed as 
cubic algebraic equations for  $\gm_{\mu\nu}$. Therefore, there can generically be 
 up to three different real solutions for  $\gm_{\mu\nu}$. 
 Since apart from $\gamma_{\mu\nu}$ equations \eqref{EG1} contain only $g_{\mu\nu}$ and $R_{\mu\nu}$,
 the solutions should be expressed  solely in terms of powers of the latter, hence 
 they  should have the structure 
  \be                           \label{solgam}
 \gm_{\mu\nu}=y_{0}\,g_{\mu\nu}+y_{1}\,R_{\mu\nu}+y_{2}\,(R^2)_{\mu\nu}
 +y_{3}\,(R^3)_{\mu\nu}. 
 \ee
 Injecting this to \eqref{EG1}, 
 eliminating the higher powers of $R_{\mu\nu}$ with the   Hamilton-Cayley relation \eqref{A}, and then setting to zero 
 the coefficients in front of $g_{\mu\nu}$, $R_{\mu\nu}$, $(R^2)_{\mu\nu}$, $(R^3)_{\mu\nu}$,
 yields a system of cubic algebraic equations for the coefficients $y_m$ (see Appendix \ref{App0}). 
 These equations will also contain 
 the parameters $\beta_A$ and the invariants \eqref{inv} of the Ricci tensor $\e_k(R^\mu_{~\nu})$,
 hence their solution will be 
 \be                     \label{bm}
 y_m=y_m(\beta_A,\e_k(R^\mu_{~\nu}),n);~~~~~~m=0,1,2,3. 
 \ee
Here the index $n=1,2,3$ counts different solutions (some of them can be complex-valued and should be rejected). 
Injecting everything  to \eqref{Mb0} yields the mass term ${\cal M}_{\mu\nu}$ with the similar 
to \eqref{solgam} structure, with $B_m=B_m(\beta_A,\e_k(R^\mu_{~\nu}))$: 
  \be                \label{mass}
 {\cal M}_{\mu\nu}=B_{0}\,g_{\mu\nu}+B_{1}\,R_{\mu\nu}+B_{2}\,(R^2)_{\mu\nu}
 +B_{3}\,(R^3)_{\mu\nu}.
 \ee

Summarizing the above discussion, the background non-linear dRGT equations 
are now fulfilled for {\it arbitrary background geometry} $g_{\mu\nu}$, 
at the expense of adjusting the reference metric. 
The linear perturbations of the background 
 are described by equations 
\be                   \label{eqs}
E_{\mu\nu}\equiv \Delta_{\mu\nu}+{\cal M}_{\mu\nu}=0. 
\ee 
Here  the kinetic term $\Delta_{\mu\nu}$ is given by \eqref{Del} while the mass term ${\cal M}_{\mu\nu}$  
is obtained by algebraically resolving \eqref{EG},\eqref{EG1} with respect to $\gm_{\mu\nu}$ and 
$\gam_{\mu\nu}$ and injecting them into \eqref{Mb0},  with 
the result of the form \eqref{mass}. The resulting mass term will depend on parameters $\beta_A$, 
in addition, there could be several different mass terms corresponding to different 
solutions \eqref{bm}.  Each mass term defines its own theory of the massive spin-2 field.

No trace of the tetrads is left: equations \eqref{eqs} describe the tensor field $X_{\mu\nu}$ evolving 
in the spacetime and their coefficients depend only on $g_{\mu\nu}$ and $R_{\mu\nu}$. 
We shall now see that these equations propagate  the 
correct number of degrees of freedom. 

\section{Constraints}

There are 16 components of $X_{\mu\nu}$ subject to 
16 equations $E_{\mu\nu}=0$. 
The essential point is that the equations imply 11 constraints which reduce the 
number of independent component of $X_{\mu\nu}$ to 5. 

\subsection{Algebraic  constraints} 
\setcounter{equation}{0}

As the operator $\Delta_{\mu\nu}$ is symmetric with respect to $\mu\leftrightarrow \nu$,
the antisymmetric part of the equations $E_{[\mu\nu]}=0$ yields 6 algebraic conditions ${\cal M}_{[\mu\nu]}=0$,
hence 
\be                     \label{MMM}
{\cal M}_{\mu\nu}={\cal M}_{\nu\mu}. 
\ee
These conditions
actually follow from the symmetry of $\gm_{\mu\nu}$ and $\gam_{\mu\nu}$.
Indeed, since these matrices are always symmetric, 
their perturbations 
should be symmetric as well, 
\be
\delta\gm_{\mu\nu}=\delta\gm_{\nu\mu},~~~~
\delta\gam_{\mu\nu}=\delta\gam_{\nu\mu},
\ee
and using  \eqref{gg} this translates to 
\be             \label{sym}
\gm^\sigma_{~\mu} X_{\sigma\nu}&=&\gm^\sigma_{~\nu} X_{\sigma\mu}\,,  \\
X_{\mu\sigma}\gam^\sigma_{~\nu}&=&X_{\nu\sigma}\gam^\sigma_{~\mu}\,,   \label{sym1}
\ee
which implies \eqref{MMM}. It is worth noting that there are only 6 independent conditions here, 
since \eqref{sym} and \eqref{sym1} follow from each other. For example, conditions \eqref{sym}
are fulfilled by setting 
\be
X_{\mu\nu}=\gam^\sigma_{~\mu} {\cal S}_{\sigma\nu}~~~~\mbox{with}~~~~~~~
{\cal S}_{\sigma\nu}={\cal S}_{\nu\sigma}
\ee
and then conditions \eqref{sym1} are fulfilled automatically. 

The latter representation suggests 
that ${\cal S}_{\mu\nu}$ could be used as  the variables instead of $X_{\mu\nu}$. However, the kinetic 
term $\Delta_{\mu\nu}$ becomes very complicated when expressed in terms of ${\cal S}_{\mu\nu}$. 
The same happens  if one uses 
$\delta\gamma_{\mu\nu}$ as the variables\footnote{This option was adopted in \cite{Guarato:2013gba},
 but  the consistency of the analysis in that paper was questioned in \cite{Bernard:2015mkk}.}. 
 The kinetic term remains simple if one 
uses  $h_{\mu\nu}=X_{\mu\nu}+X_{\nu\mu}$ to parametrize the theory  -- 
the  choice  of 
\cite{Bernard:2014bfa,Bernard:2015mkk,Bernard:2015uic}. However, the 
mass term ${\cal M}_{\mu\nu}$  then becomes extremely 
 complicated \cite{Bernard:2014bfa,Bernard:2015mkk,Bernard:2015uic}.  
 We therefore prefer using as variables 
the 16 components of $X_{\mu\nu}$ subject to 6 conditions \eqref{sym},
since both the kinetic and mass terms are then  relatively simple.

Additional   constrains on $X_{\mu\nu}$ are obtained by differentiating the equations.

\subsection{Vector constraints}
These are 
\be                       \label{C0}
{\cal C}_\nu\equiv \nabla^\mu E_{\mu\nu}=\nabla^\mu (\Delta_{\mu\nu}+{\cal M}_{\mu\nu})=0.
\ee
Using the formulas for commutators of covariant derivatives,
\be
(\nabla_\mu\nabla_\nu-\nabla_\nu\nabla_\mu)X_{\alpha\beta}
=R^\sigma_{~\alpha\nu\mu}X_{\sigma\beta}
+R^\sigma_{~\beta\nu\mu}X_{\alpha\sigma}\,,
\ee
a direct calculation yields the following result for the divergence 
of $\Delta_{\mu\nu}$ defined by \eqref{Del}:
\be              \label{20}
\nabla^\mu\Delta_{\mu\nu}&=&
G_{~\nu}^\beta (\nabla^\alpha X_{\alpha\beta}-\nabla_\beta X)\, \nn \\
&+&G^{\alpha\beta}(\nabla_\nu X_{\alpha\beta}-\nabla_\alpha X_{\beta\nu})  \nn \\
&+&X_{\alpha\beta}\nabla^\alpha G^\beta_{~\nu}.
\ee
Using the background field equations \eqref{E2}, the Einstein tensor is 
\be
G_{\mu\nu}=
-\M^{(0)}_{\mu\nu}
-\M^{(1)}_{\mu\nu}
-\M^{(2)}_{\mu\nu}
-\M^{(3)}_{\mu\nu}
\ee
where 
$
\M^{(A)}_{\mu\nu}=e^a_{~\mu}e^b_{~\nu}\M^{(A)}_{ab}
$
with $\M^{(A)}_{ab}$ given by \eqref{MM}. 
Inserting this to  \eqref{20} and \eqref{C0} yields 
\be                  \label{Cv}
{\cal C}_\nu={\cal C}^{(1)}_\nu+{\cal C}^{(2)}_\nu+{\cal C}^{(3)}_\nu
\ee
with ($A=1,2,3$)
\be              \label{C}
{\cal C}_\nu^{(A)}=&-&
\M^{(A)}_{\beta\nu}(\nabla_\alpha X^{\alpha\beta}-\nabla^\beta X)\, \nn \\
&-&\M^{(A)}_{\alpha\beta}(\nabla_\nu X^{\alpha\beta}-\nabla^\alpha X^\beta_{~\nu})  \nn \\
&-&X^{\alpha\beta}\nabla_\alpha \M^{(A)}_{\beta\nu}  \nn \\
&+&\nabla^\mu {\cal M}^{(A)}_{\mu\nu}. 
\ee
These quantities contain only the tensor $X_{\mu\nu}$ and its first derivatives. 
Let us compute them explicitly. 

\subsubsection{$\beta_1$-sector}
One has
\be
\M^{(1)}_{\mu\nu}=\beta_1([\gm]\,g_{\mu\nu}-\gm_{\mu\nu}), ~~~~~~~~~
{\cal M}^{(1)}_{\mu\nu}=\beta_1\left(
\gm_{\mu\sigma}X^\sigma_{~\nu}-\eta_{\mu\nu}\,\gm^{\alpha\beta}\,X_{\alpha\beta}\right), 
\ee 
inserting which to \eqref{C} and defining 
 \be             \label{C1b}
I^{(1)}_\nu=(\nabla_\alpha\gm_{\beta\nu}-\nabla_\nu \gm_{\alpha\beta})\,X^{\alpha\beta} 
+ (\nabla^\sigma \gm^\alpha_{~\sigma}-\nabla^\alpha[\gm])\,X_{\alpha\nu}\,
\ee
yields 
\be              \label{C1a}
\frac{1}{\beta_1}\,{\cal C}^{(1)}_\nu&=&\gm_{\nu\beta}\,(\nabla_\alpha X^{\alpha\beta}-\nabla^\beta X)
+I^{(1)}_\nu\,.
\ee

\subsubsection{$\beta_2$-sector}
Introducing 
\be                              \label{QQ}
Q_{\mu\nu}=(\gm^2)_{\mu\nu}-[\gm]\gm_{\mu\nu}
\ee
one has 
\be
\frac{1}{\beta_2}\,\M^{(2)}_{\mu\nu}
&=&Q_{\mu\nu}-\frac12\,[Q]\,g_{\mu\nu}\,, 
~~~~~\frac{1}{\beta_2}\,{\cal M}^{(2)}_{\mu\nu}=H^{\alpha\beta}_{\mu\nu}X_{\alpha\beta}\,,
\ee 
with
\be                         \label{HH}
H^{\alpha\beta}_{\mu\nu}&=&-\gm^\alpha_{~\mu}\gm^\beta_{~\nu}
+\gm_{\mu\nu}\,\gm^{\alpha\beta}-Q^\alpha_{~\mu}\delta^\beta_{~\nu}
+Q^{\alpha\beta}g_{\mu\nu}\,. 
\ee
Injecting to \eqref{C} and defining 
\be             \label{C2b}
I^{(2)}_\nu=\left(\nabla^\mu H^{\alpha\beta}_{\mu\nu}
-\nabla^\alpha(Q^\beta_{~\nu}-\frac12\,[Q]\delta^\beta_{~\nu})
\right) X_{\alpha\beta}\,
\ee
yields 
\be              \label{C2a}
\frac{1}{\beta_2}\,{\cal C}^{(2)}_\nu&=&Q_{\nu\beta}\,(\nabla^\beta X-\nabla_\alpha X^{\alpha\beta})
+\gm^\beta_{~\nu}\gm^{\alpha\sigma}(\nabla_\beta X_{\alpha\sigma}-\nabla_\sigma X_{\alpha\beta})
+I^{(2)}_\nu\,.
\ee

\subsubsection{$\beta_3$-sector}
One has 
\be
\frac{1}{\beta_3}\,\M^{(3)}_{\mu\nu}=|\gamma|\gam_{\mu\nu},~~~~~
\frac{1}{\beta_3}\,{\cal M}^{(3)}_{\mu\nu}=
|\gm|\left(X_{\mu\alpha}\gam^\alpha_{~\nu}-X\gam_{\mu\nu}\right)\,.
\ee
Injecting to \eqref{C} and defining 
\be             \label{C3b}
I^{(3)}_\nu=-X\nabla^\alpha(|\gm|\gam_{\alpha\nu})\,.
\ee
yields 
\be             \label{C3a}
\frac{1}{\beta_3}\,{\cal C}^{(3)}_\nu=|\gm|\gam_{\alpha\beta}(\nabla^\alpha X^\beta_{~\nu}
-\nabla_\nu X^{\alpha\beta})
+I^{(3)}_\nu\,.
\ee

\subsubsection{Vector constraints and the massless limit} 
Adding up the quantities ${\cal C}^{(A)}_\nu$ in \eqref{C1a},\eqref{C2a},\eqref{C3a} 
yields 
\be                       \label{Cv1}
0={\cal C}_\nu\equiv \nabla^\mu E_{\mu\nu}&=&{\cal C}^{(1)}_\nu+{\cal C}^{(2)}_\nu+{\cal C}^{(3)}_\nu \nn \\
=&&\beta_1 \,
\gm_{~\nu}^\beta\,(\nabla^\alpha X_{\alpha\beta}-\nabla_\beta X)  \nn \\
&+&\beta_2\,\gm_{~\nu}^\beta\, \{\,
(\gm^\sigma_{~\beta}-[\gm]\delta^\sigma_{~\beta}) (\nabla_\sigma X-\nabla^\alpha X_{\alpha\sigma})
+\gm^{\alpha\sigma}(\nabla_\beta X_{\alpha\sigma}-\nabla_\sigma X_{\alpha\beta})\,\} \nn \\
&+&\beta_3\,|\gm|\gam_{\alpha\beta}(\nabla^\alpha X^\beta_{~\nu}-\nabla_\nu X^{\alpha\beta})
)  \nn \\
&+&
\beta_1\left\{
(\nabla_\alpha\gm_{\beta\nu}-\nabla_\nu \gm_{\alpha\beta})\,X^{\alpha\beta} 
+ (\nabla^\sigma \gm^\alpha_{~\sigma}-\nabla^\alpha[\gm])\,X_{\alpha\nu}\,
\right\}  \nn   \\
&+&\beta_2 
\left(\nabla^\mu H^{\alpha\beta}_{\mu\nu}-\nabla^\alpha(Q^\beta_{~\nu}-\frac12\,[Q]\delta^\beta_{~\nu})
\right) X_{\alpha\beta} \nn  \\
&-&\beta_3\,X\nabla^\alpha(|\gm|\gam_{\alpha\nu})\,. 
\ee
These quantities vanish on-shell, where $E_{\mu\nu}=0$, 
which yields 
 4 relations between $X_{\alpha\beta}$ and $\nabla_\sigma X_{\alpha\beta}$, 
hence 4 constraints for the initial data. Together with the 6 algebraic constraints \eqref{sym}, 
this reduces the number of DoF contained in $X_{\alpha\beta}$ to $16-6-4=6$. 

It is also possible  that ${\cal C}_\nu$ may vanish off-shell, for any $X_{\mu\nu}$. 
One has 
\be
{\cal C}_\nu= 2{\cal A}^{\alpha\beta\sigma}_\nu \nabla_\sigma X_{\alpha\beta}+
{\cal B}^{\alpha\beta}_\nu X_{\alpha\beta}  
\ee
with 
\begingroup
\addtolength{\jot}{1.em}
\be 
{\cal A}^{\alpha\beta\sigma}_\nu &=&  \beta_{1} g^{\alpha[\sigma } \gamma^{\beta]}_{\nu} 
+ \beta_{2} \big ( g^{\alpha [\beta} Q^{\sigma]}_{\nu} + \gamma^{\alpha [\beta} \gamma^{\sigma]}_{\nu}
 \big )
+ \beta_{3}|\gm|  \delta^{[\beta}_{\nu} \gam^{\sigma] \alpha},  \nn \\
{\cal B}^{\alpha\beta}_\nu&=&  \beta_{1} \big [ \nabla^{\alpha} \gamma^{\beta}_{\nu} 
- \nabla_{\nu} \gamma^{\alpha \beta} + \delta^{\beta}_{\nu} ( \nabla^{\sigma} \gamma^{\alpha}_{\sigma} 
- \nabla^{\alpha} [\gamma] ) \big ]  \nn \\
&+& \beta_{2} \big [ \nabla^{\mu} H^{\alpha \beta}_{\, \mu \nu} - \nabla^{\alpha}(Q^{\beta}_{\, \nu} - \frac{1}{2} [Q] \delta^{\beta}_{\nu}) \big ] 
 - \beta_{3}\, g^{\alpha \beta} \nabla^{\sigma}( |\gm|\gam_{\sigma \nu})  ,
\ee
\endgroup
hence ${\cal C}_\nu$ will vanish identically if the background is such that 
${\cal A}^{\alpha\beta\sigma}_\nu =0$ and ${\cal B}^{\alpha\beta}_\nu=0$. 
The constraints ${\cal C}_\nu$  generate
in this case  gauge transformations and one should count them  twice. 
As a result, the number of degrees of freedom reduces to $6-4=2$, which corresponds to 
two polarizations of massless spin-2 particles. 
Therefore, the conditions ${\cal A}^{\alpha\beta\sigma}_\nu =0$ and ${\cal B}^{\alpha\beta}_\nu=0$
describe  the massless limit of the theory. 

However, unless for $\beta_1=\beta_2=\beta_3$, the massless limit is possible only for 
special backgrounds. It seems that for generic 
 $\beta_A$ the conditions ${\cal A}^{\alpha\beta\sigma}_\nu =0$ and ${\cal B}^{\alpha\beta}_\nu=0$ hold 
 if only 
$\gm_{\mu\nu}=0$, in which case the background Einstein equations \eqref{EG1} reduce to 
\be
G_{\mu\nu}+\beta_0\,g_{\mu\nu}=0,
\ee
hence the background is an Einstein space. Therefore, the  massive spin-2 field can become 
massless only in Einstein spaces. For any other background it always carries 
5 (as we shall now see) degrees of freedom.

\subsection{Scalar constraint}
Let us return to the quantities ${\cal C}^{(A)}_\nu$ 
computed in \eqref{C1a},\eqref{C2a},\eqref{C3a}
and differentiate them. 
\subsubsection{$\beta_1$-sector}
We notice that the part of ${\cal C}^{(1)}_\nu$ containing the derivatives of $X_{\mu\nu}$ 
is proportional to the matrix $\gm_{\nu\beta}$ (see \eqref{C1a}). Therefore, multiplying 
by the inverse matrix $\gam^{\sigma\nu}$ and acting with $\nabla_\sigma$ yields 
\be              
\frac{1}{\beta_1}\,\nabla_\sigma(\gam^{\sigma\nu}{\cal C}^{(1)}_\nu)
&=&\nabla_\sigma\nabla_\alpha X^{\alpha\sigma}-\Box X
+\nabla_\sigma(\gam^{\sigma\nu}I^{(1)}_\nu),
\ee
where $\nabla_\sigma\nabla_\alpha X^{\alpha\sigma}=\nabla_\alpha\nabla_\sigma X^{\alpha\sigma}$. 
On the other hand, taking the trace of the equations gives
\be
\frac12 E^\mu_{~\mu}=\Box X-\nabla^\alpha\nabla^\beta X_{\alpha\beta}+R^{\alpha\beta}X_{\alpha\beta} 
+\frac12\,{\cal M}^\mu_{~\mu}.
\ee
Therefore, the combination 
\be                \label{b11}
{\cal C}^{(1)}_5\equiv \nabla_\sigma(\gam^{\sigma\nu}{\cal C}^{(1)}_\nu)+\frac{\beta_1}{2}E^\mu_{~\mu}=
\beta_1\left(
\nabla_\sigma(\gam^{\sigma\nu}I^{(1)}_\nu)+R^{\alpha\beta}X_{\alpha\beta} 
+\frac12\,{\cal M}^\mu_{~\mu}
\right)
\ee
does not contain second derivatives of $X_{\mu\nu}$. 

\subsubsection{$\beta_2$-sector}
The part of ${\cal C}^{(2)}_\nu$ containing the derivatives of $X_{\mu\nu}$ in  \eqref{C2a}
is also proportional to the matrix $\gm_{\nu\beta}$. 
This yields 
\be              
\frac{1}{\beta_2}\,\nabla_\mu(\gam^{\mu\nu}{\cal C}^{(2)}_\nu)
&=&\nabla_\mu\left\{
(\gamma^\mu_{~\beta}-\gamma\delta^\mu_{~\beta})(\nabla^\beta X-\nabla_\alpha X^{\alpha\beta})
+\gamma^{\alpha\sigma}(\nabla^\mu X_{\alpha\sigma}-\nabla_\sigma X_\alpha^{~\mu})
\right\}   \nn  \\
&+&\nabla_\mu(\gam^{\mu\nu}I^{(2)}_\nu)  \\ 
&=& {\cal D}+J^{(2)}
\ee
with 
\be
{\cal D}&=&(\gm^{\alpha\beta}\Box+\gm\nabla^\alpha\nabla^\beta)X_{\alpha\beta}+
(\gm^{\alpha\beta}\nabla_\alpha\nabla_\beta-\gamma\Box)X             \nn
-\gm^{\mu\nu}\nabla_\mu\nabla_\sigma (X^\sigma_{~\nu}+X_\nu^{~\sigma}) \nn \\
J^{(2)}&=&
(\nabla^\beta X-\nabla_\alpha X^{\alpha\beta})\nabla_\mu(\gm^\mu_{~\beta}-\gm\delta^\mu_{~\beta})
+(\nabla^\mu X_{\alpha\sigma}-\nabla_\sigma X_\alpha^{~\mu})\nabla_\mu\gm^{\alpha\sigma}
   \nn  \\
   &+&\nabla_\mu(\gam^{\mu\nu}I^{(2)}_\nu) +(R^\nu_{~\alpha\mu\sigma}\gm^{\alpha\sigma}
   -R^{\alpha\nu}\gm_{\mu\alpha})X^\mu_{~\nu}\,.
\ee
On the other hand, one has 
\be
\gm^{\mu\nu}\Delta_{\mu\nu}&=&-{\cal D}
+\gm R^{\alpha\beta}X_{\alpha\beta}-2\gm^{\mu\nu}R_{\sigma\mu}X^\sigma_{~\nu}\,.
\ee
As a result, the sum
\be                         \label{b22}
{\cal C}^{(2)}_5&\equiv &\nabla_\sigma(\gam^{\sigma\nu}{\cal C}^{(2)}_\nu)+\beta_2 \gm^{\mu\nu}E_{\mu\nu}  \\
&=&
\beta_2
\left(J^{(2)}+
\gm R^{\alpha\beta}X_{\alpha\beta}-2\gm^{\mu\nu}R_{\sigma\mu}X^\sigma_{~\nu}+\gm^{\mu\nu}{\cal M}_{\mu\nu}
\right)  \nn
\ee
does not contain second derivatives of $X_{\mu\nu}$. 

\subsubsection{$\beta_3$-sector}
Using \eqref{C3a} yields 
\be
\frac{1}{\beta_3}\,\gam^{\mu\nu}{\cal C}^{(3)}_\nu=
|\gm|(\gam^{\mu\beta}\gam^{\nu\alpha}-\gam^{\mu\nu}\gam^{\alpha\beta})
\nabla_\nu X_{\alpha\beta}-X\gam^{\mu\nu}\nabla^\alpha(|\gm|\gam_{\alpha\nu})
\ee
hence 
\be                        \label{b3}
\frac{1}{\beta_3}\,\nabla_\mu(\gam^{\mu\nu}{\cal C}^{(3)}_\nu)=
|\gm|(\gam^{\mu\beta}\gam^{\nu\alpha}-\gam^{\mu\nu}\gam^{\alpha\beta})
\nabla_\mu\nabla_\nu X_{\alpha\beta}
+J^{(3)}
\ee
with 
\be
J^{(3)}=
\nabla_\nu X_{\alpha\beta}
\nabla_\mu\left\{|\gm|(\gam^{\mu\beta}\gam^{\nu\alpha}-\gam^{\mu\nu}\gam^{\alpha\beta})
\right\}-\nabla_\mu\left\{X\gam^{\mu\nu}\nabla^\alpha(|\gm|\gam_{\alpha\nu})\right\}. 
\ee
Now, the right hand side in \eqref{b3}  does contain the second derivatives of $X_{\mu\nu}$, but
the second {\it time} derivatives are contained 
only in 
\be                  \label{TTT}
|\gm|(\gam^{0\alpha}\gam^{0\beta}-\gam^{00}\gam^{\alpha\beta})\,
\ddot{X}_{\alpha\beta}=
|\gm|(\gam^{0i}\gam^{0k}-\gam^{00}\gam^{ik})\,
\ddot{X}_{ik}\,.
\ee
The second derivatives $\ddot{X}_{ik}$ can be expressed from the 
field equations. 
The definition of $\Delta_{\mu\nu}$ in \eqref{Del} implies that 
\be
\Delta_{ik}=-g^{00}\ddot{X}_{(ik)}+g_{ik} \,g^{00} \,\hh^{nm} \ddot{X}_{nm}+\ldots 
\ee
where $\hh^{ik}=g^{ik}-g^{0i}g^{0k}/g^{00}$ is the inverse of $g_{ik}$ and the dots denote terms 
not containing $\ddot{X}_{\mu\nu}$. Inverting this relation yields 
\be
\ddot{X}_{(ik)}&=&\frac{1}{g^{00}}\left(\frac{1}{2}\,{g_{ik}}\,\hh^{nm}\Delta_{nm}-\Delta_{ik}  \right)+\ldots \nn \\
 &=&\frac{1}{g^{00}}\left(\frac{1}{2}\,{g_{ik}}\,\hh^{nm} E_{nm}-E_{ik}  \right)+\ldots
\ee
Therefore,  the combination 
\be
\frac{|\gm|}{g^{00}}(\gam^{0i}\gam^{0k}-\gam^{00}\gam^{ik})\left(\frac{1}{2}\,{g_{ik}}\,\hh^{nm} E_{nm}-E_{ik}  \right)
~~~~~~ \\
=\frac{|\gm|}{
g^{00}}(
\gam^{0\alpha}\gam^{0\beta}-
\gam^{00}\gam^{\alpha\beta}
)\left(\frac{1}{2}\,{g_{\alpha\beta}}\,\hh^{nm} E_{nm}-E_{\alpha\beta}  \right) \nn
\ee
has precisely the same second time derivatives as \eqref{TTT}. Noting finally that 
\be
\hh^{nm} E_{nm}&=&\frac{1}{g^{00}}(g^{00}g^{mn}-g^{0m}g^{0n}) E_{mn}   \\
&=&\frac{1}{g^{00}}(g^{00}g^{\mu\nu}-g^{0\mu}g^{0\nu}) E_{\mu\nu}
= E^\alpha_{~\alpha}-\frac{1}{g^{00}}\, E^{00}  \nn
\ee
it follows that the quantity  
\be                        \label{b33}
{\cal C}^{(3)}_5\equiv 
\frac{1}{\beta_3}\,\nabla_\mu(\gam^{\mu\nu}{\cal C}^{(3)}_\nu)+
\frac{|\gm|}{g^{00}}(\gam^{0\alpha}\gam^{0\beta}-\gam^{00}\gam^{\alpha\beta})\,
(
E_{\alpha\beta}-\frac12\,g_{\alpha\beta}(E^\sigma_{~\sigma}-\frac{1}{g^{00}}\,E^{00}  )
)
\ee
does not contain $\ddot{X}_{\mu\nu}$. This quantity is not generally covariant 
and depends on the time choice, but for any such a choice the second derivatives with 
respect to the corresponding time  coordinate cancel each other. 

Summing up the above expressions \eqref{b11},\eqref{b22},\eqref{b33} for ${\cal C}^{(A)}_5$ we obtain 
\be            \label{C5}
0={\cal C}_5&\equiv &\beta_1{\cal C}^{(1)}_5+\beta_2{\cal C}^{(2)}_5+\beta_3{\cal C}^{(3)}_5  \nn \\
&=& \nabla_\mu(\gam^{\mu\nu}\nabla^\sigma E_{\sigma\nu})
+\frac{\beta_1}{2}\,E^\alpha_{~\alpha}+\beta_2\gm^{\mu\nu}E_{\mu\nu} \nn \\
&+&\beta_3\left(
\frac{|\gm|}{g^{00}}(\gam^{0\alpha}\gam^{0\beta}-\gam^{00}\gam^{\alpha\beta})\,
\left(
E_{\alpha\beta}-\frac12\,g_{\alpha\beta}(E^\sigma_{~\sigma}-\frac{1}{g^{00}}\,E^{00}  )
\right)
\right).
\ee
This does not contain $\ddot{X}_{\mu\nu}$ and vanishes on-shell.  Hence this is an additional  constraint 
on the initial data that reduces the number of  DoF from 6 to 5. 
It is remarkable that in our theory this constraint can be expressed  in a simple 
and covariant (for $\beta_3=0$) form. 

This constraint can also be rewritten as 
\begin{equation} \label{C5a}
\mathcal{C}_{5} = \mathfrak{A}^{\lambda \sigma \alpha \beta} \nabla_{\lambda} \nabla_{\sigma} X_{\alpha \beta} 
+ \mathfrak{B}^{\sigma \alpha \beta} \nabla_ {\sigma} X_{\alpha \beta} + \mathfrak{C}^{\alpha \beta} X_{\alpha \beta}
\end{equation}
where the coefficients $\mathfrak{A}^{\lambda \sigma \alpha \beta}$, $\mathfrak{B}^{\sigma \alpha \beta}$, and 
$\mathfrak{C}^{\alpha \beta}$ are given  in  Appendix~\ref{AppA}. 
If all these coefficients vanish then the background is {\it partially massless} (PM) since the constraint then 
generates gauge transformations and there remain only 4 dynamical DoF. 
The PM backgrounds can be Einstein spaces, but it seems this is not the only possibility 
\cite{Bernard:2017tcg}\footnote{We have not studied the PM backgrounds in our theory. 
Ref.\cite{Bernard:2017tcg} presents some PM solutions which are not Einstein spaces for the case where $\beta_3=0$.}. 

As the final remark, we notice that our expression for the scalar contraint can be applied also 
within in the original non-linear dRGT theory. In fact, the existence of the scalar  constraint  in this theory 
can be shown within the Hamiltonian approach, but this  
requires tedious calculations  \cite{Hassan:2011ea}. 
However, since the background dRGT equations ${\bf E}_{\mu\nu}=0$ in \eqref{EG} are linear in the 
second derivatives, the latter are exactly the same as in 
the linearized equations 
$E_{\mu\nu}=0$ expressed by \eqref{eqs}. Therefore, simply  replacing  in \eqref{C5} $E_{\mu\nu}$ by 
${\bf E}_{\mu\nu}$ yields  the expression not containing the second (time) derivatives of the metric $g_{\mu\nu}$, 
\be            \label{C55}
0=\bm{C}_5
&\equiv & \nabla_\mu(\gam^{\mu\nu}\nabla^\sigma {\bf E}_{\sigma\nu})
+\frac{\beta_1}{2}\,{\bf E}^\alpha_{~\alpha}+\beta_2\gm^{\mu\nu}{\bf E}_{\mu\nu} \nn \\
&+&\beta_3\left(
\frac{|\gm|}{g^{00}}(\gam^{0\alpha}\gam^{0\beta}-\gam^{00}\gam^{\alpha\beta})\,
\left(
{\bf E}_{\alpha\beta}-\frac12\,g_{\alpha\beta}({\bf E}^\sigma_{~\sigma}-\frac{1}{g^{00}}\,{\bf E}^{00}  )
\right)
\right).
\ee
This is the scalar constraint in the dRGT theory.

\section{Two special models} 
\setcounter{equation}{0}

Summarizing the above discussion, massive spin-2 particles 
 in curved space can be described by a non-symmetric tensor 
 $X_{\mu\nu}$ that 
fulfills equations \eqref{eqs}
where  the kinetic term $\Delta_{\mu\nu}$ and the mass term ${\cal M}_{\mu\nu}$ are defined by 
\eqref{Del} and by \eqref{Mb0}. 
 The equations imply 6 algebraic conditions \eqref{sym} and five differential constraint \eqref{Cv1} and \eqref{C5}
 which reduce the number of independent components of 
$X_{\mu\nu}$  from 16 to 5. This matches the number of polarisations of massive spin-2 particles. 

The background geometry can be arbitrary. 
The mass term ${\cal M}_{\mu\nu}$ depends on it 
via matrices $\gm_{\mu\nu}$ and $\gam_{\mu\nu}$ algebraically 
 related to the background metric and $R_{\mu\nu}$ by conditions \eqref{EG} or \eqref{EG1}. 
 The dependence of ${\cal M}_{\mu\nu}$ on $R_{\mu\nu}$ 
 is in general non-linear, but  it becomes linear in two special cases that we call model~I and model~II.
 These two models will be discussed 
 in the rest of the text. 
 
 \subsection{Model I}
 Setting in \eqref{EG} $\beta_2=\beta_3=0$ one obtains 
\be                         \label{EG2} 
G_{\mu\nu}+\beta_0\,\eta_{\mu\nu}+\beta_1([\gm]\,g_{\mu\nu}-\gm_{\mu\nu})
=0, 
\ee
from where 
\be
\beta_1\gm_{\mu\nu}=R_{\mu\nu}-\left(\frac{R}{6}+\frac{\beta_0}{3}\right)g_{\mu\nu}\equiv \gmm_{\mu\nu}.
\ee
Injecting this to \eqref{Mb} yields the mass term 
\be                     \label{2}
{\cal M}_{\mu\nu}&= &\gmm_{\mu\alpha} X^\alpha_{~\nu}
-g_{\mu\nu}\,\gmm_{\alpha\beta}X^{\alpha\beta}  
\ee
with 
\be                       \label{gamI}
\gmm_{\mu\nu}&=&R_{\mu\nu}+\left(\muu -\frac{R}{6}\right)g_{\mu\nu}\
\ee
where 
\be
M^2=-{\beta_0}/{3}
\ee
plays the role of the FP mass. 
Notice that the dependence on $\beta_1$ has gone. 
The field equations are 
$E_{\mu\nu}\equiv \Delta_{\mu\nu}+{\cal M}_{\mu\nu}=0$ with  $\Delta_{\mu\nu}$ given by \eqref{Del}. 

It is worth checking again the constraints. 
 The 
asymmetric part, $E_{[\mu\nu]}=0$, yields 6 algebraic conditions  
\be                       \label{3} 
\gmm_{\mu\alpha} X^\alpha_{~\nu}=\gmm_{\nu\alpha} X^\alpha_{~\mu}\,.
\ee
Taking the divergence of $E_{\mu\nu}$ 
and defining 
\be
{\cal I}_\nu= X^{\alpha\beta}(\nabla_\alpha G_{\beta\nu}
-\nabla_\nu\gmm_{\alpha\beta})+\nabla^\mu\gmm_{\mu\alpha}X^\alpha_{~\nu}
\ee
yields  (see Appendix \ref{AppB}) four vector constraints, 
\be                  \label{vI}
0={\cal C}_\nu\equiv \nabla^\mu E_{\mu\nu}=\gmm_{\nu\rho}(\nabla_\sigma X^{\sigma\rho}-\nabla^\rho X)
+{\cal I}_\nu. 
\ee
Multiplying this by the inverse  $(\gmm^{-1})^{\rho\nu}$ of $\gmm_{\rho\nu}$
and 
taking the divergence again yields (see Appendix \ref{AppB}) the fifth constraint, 
\be                \label{sI}
0={\cal C}_5&\equiv& \nabla_\rho ((\gmm^{-1})^{\rho\nu} \nabla^\mu E_{\mu\nu})+\frac12 E^\mu_{~\mu}  \nn \\
&=&-\frac32\muu X-\frac12 G^{\mu\nu}X_{\mu\nu}
+\nabla_\rho((\gmm^{-1})^{\rho\nu}  {\cal I}^\nu).~~~
\ee

\subsection{Model II}
Getting back to dRGT equations \eqref{EG} for generic $\beta_A$ and setting $\beta_1=\beta_2=0$ yields 
 \be                         \label{EG3} 
G_{\mu\nu}+\beta_0\,g_{\mu\nu}
+\beta_3\,|\gm|\gam_{\mu\nu}=0, 
\ee
hence 
\be
-\beta_3\,|\gm|\,\gam_{\mu\nu}=G_{\mu\nu}+\beta_0\,g_{\mu\nu}\equiv \gmm_{\mu\nu}.
\ee
Injecting this to  \eqref{Mb} yields the mass term
\be                   \label{2a}
{\cal M}_{\mu\nu}&=& -X_\mu^{~\alpha} \gmm_{\alpha\nu} 
+X\gmm_{\mu\nu}\,,  
\ee
where
\be                      \label{gamII}
\gmm_{\mu\nu}&=&R_{\mu\nu}-\left(\muu +\frac{R}{2}\right)g_{\mu\nu}\,
\ee
with the FP mass 
\be
M^2=-\beta_0.
\ee
Injecting \eqref{2a} to \eqref{eqs} yields the equations. Taking 
again
 the 
asymmetric part of the equations, $E_{[\mu\nu]}=0$, yields 6 algebraic conditions  
\be                       \label{3a} 
X_\mu^{~\alpha} \gmm_{\alpha\nu} =X_\nu^{~\alpha} \gmm_{\alpha\mu} \,, 
\ee
while  taking the divergence of $E_{\mu\nu}$ yields (see Appendix \ref{AppB}) the vector constraints 
\be                          \label{vII}
0={\cal C}_\nu\equiv\nabla^\mu E_{\mu\nu}=
\gmm^{\alpha\beta}(\nabla_\nu\, X_{\alpha\beta}-\nabla_\alpha\, X_{\beta\nu}). 
\ee
Multiplying this by $\gmm^{\rho\nu}=g^{\rho\alpha}g^{\nu\beta}\gmm_{\alpha\beta}$ 
(not to be confused with the  the inverse $(\gmm^{-1})^{\rho\nu}$), 
taking the divergence 
and combining with the equations  yields (see Appendix \ref{AppB})
\be                         \label{sII}
0={\cal C}_5\equiv \nabla_\rho (\gmm^{\rho\nu}{\cal C}_\nu)+\frac{1}{2g^{00}}\,
(\gmm^{00}\gmm^{\alpha\beta}-\gmm^{0\alpha}\gmm^{0\beta})
\Big (
E_{\alpha\beta}-\frac12\,g_{\alpha\beta}(E^\sigma_{~\sigma}-\frac{1}{g^{00}}\,E^{00}  )
\Big ).~~~
\ee
This does not contain $\ddot{X}_{\mu\nu}$
hence this is a constraint. 

\subsection{Action} 

Equations $E_{\mu\nu}=\Delta_{\mu\nu}+{\cal M}_{\mu\nu}=0$ with ${\cal M}_{\mu\nu}$ 
given either by \eqref{2} (model I) or by \eqref{2a} (model II) can be obtained by varying the action 
\be                        \label{act}
I=\frac12\int X^{\nu\mu}E_{\mu\nu}\sqrt{-g}\,d^4x\equiv \int L \,\sqrt{-g}\,d^4x
\ee
(notice the order of indices). 
One can split the Lagrangian into the kinetic term and the potential term,
\be                          \label{Lag}
L=L_{(2)}+L_{(0)}, 
\ee
where,  
after integrating by parts, the kinetic term is 
\be                          \label{Lag1}
L_{(2)}=&-&\frac14\, \nabla^\sigma \h^{\mu\nu}\nabla_\mu \h_{\nu\sigma}
+\frac18\,\nabla^\alpha \h^{\mu\nu} \nabla_\alpha \h_{\mu\nu}   \nn \\
&+&\frac14 \nabla^\alpha \h \nabla^\beta \h_{\alpha\beta}-\frac18\, \nabla_\alpha \h\nabla^\alpha \h
\ee
with $\h_{\mu\nu}=X_{\mu\nu}+X_{\nu\mu}$ and $\h=\h^\alpha_{~\alpha}$. 
The potential term in model~I is 
\be                          \label{Lag2}
L_{(0)}=&-&\frac12\,X^{\mu\nu}R^\sigma_{~\mu}X_{\sigma\nu} \nn \\
&+&\frac12\,(M^2-\frac{R}{6})(X_{\mu\nu}X^{\nu\mu}-X^2),
\ee
and in model II
\be                          \label{Lag3}
L_{(0)}=&-&\frac12\,X^{\mu\nu}R^\sigma_{~\mu}X_{\sigma\nu} 
-\frac12\,X^{\mu\nu}R^\sigma_{~\nu}X_{\sigma\mu}   \nn \\
&-&\frac12X^{\mu\nu}X_{\nu\alpha}R^\alpha_{~\mu}+XR_{\mu\nu}X^{\mu\nu} \nn \\
&+&\frac12\,(M^2+\frac{R}{2})(X_{\mu\nu}X^{\nu\mu}-X^2);
\ee
the order of indices being important. One can directly check that 
varying the action with respect to $X_{\mu\nu}$ yields the field equations, 
\be
\delta I=\int E_{\nu\mu}\, \delta X^{\mu\nu}\,\sqrt{-g}\,d^4x.
\ee
Varying with respect to the metric gives the energy-momentum tensor, 
\be
\delta I=-\frac12 \int T_{\mu\nu}\,\delta g^{\mu\nu}\,\sqrt{-g}\,d^4 x. 
\ee

\section{Massive spin-2 field in Einstein spaces}

We shall now study the equations in models I and II
for specific backgrounds. To begin with, we show that if the background is  an Einstein space, 
hence 
$R_{\mu\nu}=\Lambda g_{\mu\nu}$, then the 
equations $E_{\mu\nu}\equiv \Delta_{\mu\nu}+{\cal M}_{\mu\nu}=0$ reproduce 
the standard description of massive gravitons. 
Indeed,  then in both models 
the tensor $\gmm_{\mu\nu}$ becomes proportional to the metric and the conditions \eqref{3},\eqref{3a}  yield 
$X_{\mu\nu}=X_{\nu\mu}$. Equations reduce to 
\be             \label{4}
E_{\mu\nu}\equiv \Delta_{\mu\nu}+M_{\rm H}^2(X_{\mu\nu}-[X]g_{\mu\nu})=0
\ee
with 
\be                    \label{Dell}
\Delta_{\mu\nu}&=&\nabla^\sigma\nabla_\mu X_{\sigma\nu} 
 +\nabla^\sigma\nabla_\nu X_{\sigma\mu} 
-\Box {X}_{\mu\nu}-\nabla_\mu\nabla_\nu{[X]}  \nn \\
 &+&g_{\mu\nu}\left(\Box [X]-\nabla^\alpha\nabla^\beta X_{\alpha\beta}+\Lambda [X]
\right) -2\Lambda X_{\mu\nu}
\ee
where  the Higuchi mass \cite{Higuchi} is 
\be
\mbox{model I:}~~~M_{\rm H}^2={\Lambda}/{3}+\muu;~~~~~~~~~~~~
\mbox{model II:}~~~M_{\rm H}^2={\Lambda}+\muu\,.
\ee
The operator $\Delta_{\mu\nu}$ in \eqref{Dell} is divergence free, $\nabla^\mu\Delta_{\mu\nu}=0$ 
(see Appendix \ref{AppB}),
and is invariant under 
\be
X_{\mu\nu}\to X_{\mu\nu}+\nabla_{\mu} \xi_{\nu}+\nabla_{\nu} \xi_{\mu}. 
\ee
For  $M_{\rm H}=0$ this becomes the gauge symmetry of the equations which 
describe   in this case  
massless gravitons with two polarizations. 

If $M_{\rm H}\neq 0$ then, taking the divergence of \eqref{4}, yields 
four constraints $\nabla^\mu X_{\mu\nu}=\nabla_\nu [X]$.  Using them  reduces equations
\eqref{4}  to 
\be                \label{5}
&-&\Box X_{\mu\nu}+\nabla_\mu\nabla_\nu [X]-2R_{\mu\alpha\nu\beta}X^{\alpha\beta} 
+\Lambda [X]g_{\mu\nu}
+M^2_{\rm H}(X_{\mu\nu}-[X]g_{\mu\nu})=0.
\ee
The trace of these yields 
$
(2\Lambda-3M_{\rm H}^2)[X]=0
$
hence, unless for $M^2_{\rm H}=2\Lambda/3$, one has $[X]=0$.
This  is the fifth constraint reducing the 
number of degrees of freedom to five. 

In the exceptional case where $M^2_{\rm H}=2\Lambda/3\equiv M^2_{\rm PM}$ the trace $[X]$ does not vanish,
but 
equations \eqref{5} 
are then invariant under 
\be
X_{\mu\nu}\to X_{\mu\nu}+\nabla_\mu\nabla_\nu\Omega+\frac{\Lambda}{3}\,g_{\mu\nu}\Omega. 
\ee
Due to this gauge symmetry 
there remain only four degrees of freedom. This corresponds to 
the partially massless (PM) case \cite{Higuchi}.
 
We conclude that  our theory successfully reproduces the 
standard properties of massive gravitons in Einstein spaces.

\section{Massive spin-2 field in expanding universe}

Let us now assume the background metric $g_{\mu\nu}$ to be a solution 
of the Einstein equations 
\be                  \label{EQS}
M_{\rm Pl}^2\, G_{\mu\nu}=T^{\rm (m)}_{\mu\nu}\,,
\ee
where $T^{\rm (m)}_{\mu\nu}$ is the energy-momentum tensor of some matter source. 
Choosing  the geometry to be homogeneous and isotropic of the spatially flat FLRW type, 
\be                \label{cosm}
g_{\mu\nu}dx^\mu dx^\nu=-dt^2+a^2(t)d{\bf x}^2,
\ee
while the matter to be a perfect fluid, 
$T^{\rm (m) \mu}_{~~~~~\nu}={\rm diag}[-\bm{\rho}(t),\bm{p}(t),\bm{p}(t),\bm{p}(t)]$, 
the Einstein equations  \eqref{EQS} reduce to 
\be
3\,H^2=\frac{\bm{\rho}}{M^2_{\rm Pl}}\equiv \rho,
~~~~\dot{H}=-\frac{\bm{\rho+p}}{2M^2_{\rm Pl}}\equiv -\frac{\rho+p}{2},
\ee
where $H=\dot{a}/a$ is the Hubble parameter. 

We wish to construct the { general} solution of 
$E_{\mu\nu}\equiv \Delta_{\mu\nu}+{\cal M}_{\mu\nu}=0$ 
with $\Delta_{\mu\nu}$ given by \eqref{Del} and ${\cal M}_{\mu\nu}$ defined 
either  by \eqref{2} or by \eqref{2a}
on the cosmological background \eqref{cosm}.
The general solution for  $X_{\mu\nu}$ can be represented in this case as 
\be                     \label{XX0}
X_{\mu\nu}(t,{\bf x})=\sum_{\bf k}X_{\mu\nu}(t,{\bf k})e^{i{\bf kx}}
\ee where the Fourier amplitude 
splits into the sum of the tensor, vector, and scalar harmonics:
\be                    \label{XX1}
X_{\mu\nu}(t,{\bf k})=X^{(2)}_{\mu\nu}+X^{(1)}_{\mu\nu}+X^{(0)}_{\mu\nu}. 
\ee
Since the spatial part of the background 
Ricci tensor is proportional to the unit matrix, $R_{ik}\sim \delta_{ik}$, the algebraic 
constraints \eqref{3},\eqref{3a}  imply that $X_{ik}=X_{ki}$ hence $X_{\mu\nu}$ has in this case only 13 
independent components. Assuming the spatial momentum $\bf{k}$ to be directed along 
the third axis, ${\bf k}=(0,0,{\rm k})$, the harmonics can be parameterized as 
\be                          \label{XX}
X^{(2)}_{\mu\nu}&=&
\begin{bmatrix}
    0       & 0 & 0 & 0 \\
    0       & {\rm D_+} & {\rm D_-} & 0 \\
    0       & {\rm D_-} & {\rm -D_+} & 0 \\
       0       & 0 & 0 & 0 \\
 \end{bmatrix},~
 X^{(1)}_{\mu\nu}=
\begin{bmatrix}
    0       & W_+^+ & W_-^+ & 0 \\
    W^-_+       & 0 & 0 & i{\rm k}{\rm V_+} \\
    W^-_-       & 0 & 0 & i{\rm k}{\rm V_-} \\
    0       & i{\rm k}{\rm V_+}& i{\rm k}{\rm V_-} & 0 \\
 \end{bmatrix},~     \nn \\
 X^{(0)}_{\mu\nu}&=&
\begin{bmatrix}
    S^+_+       & 0 & 0 & i{\rm k}S^+_- \\
    0       & S^-_- & 0 & 0 \\
    0       & 0 & S^-_- & 0 \\
   i{\rm k} S^-_+       & 0 & 0 & S^-_--{\rm k}^2{\rm S} \\
 \end{bmatrix},~~
\ee
where ${\rm D_\pm}$, ${\rm V}_{\pm}$, ${\rm S}$, $W^\pm_\pm$, $S^\pm_\pm$  are 
functions of time. Injecting everything to $E_{\mu\nu}=0$, 
the equations split into three independent groups -- one for the tensor modes 
$X^{(2)}_{\mu\nu}$, one for the vector modes $X^{(1)}_{\mu\nu}$, and one for  scalar modes 
$X^{(0)}_{\mu\nu}$. 

In the tensor sector everything reduces to two separate  second order equations 
for ${\rm D}_+$ and ${\rm D}_-$ describing the two tensor polarizations. 
In the vector sector  the four amplitudes $W^\pm_\pm$ can be expressed by virtue of the field  equations 
 (see Appendix \ref{AppC}) 
in terms of 
${\rm V}_{+}$ and ${\rm V}_{-}$
which  fulfill two separate second order equations describing the two vector polarizations. 

Most importantly, one finds that in the scalar sector 
the four $S^\pm_\pm$ can be expressed  (see Appendix \ref{AppC})  
in terms of one single amplitude S that fulfills a separate second order 
equation. Therefore, there is only one scalar polarization, hence there are altogether 5 DoF. 

Injecting everything into the action \eqref{act}, it splits into the sum of five terms of the form 
\be                    \label{act1}
\left.\left.\int \right( K_{(s)} \dot{Y}^2-U_{(s)}Y^2\right) a^3\,dt;~~~~~~s=0,1,2. 
\ee
For the tensor modes one has 
$Y=D_{+}$ or $Y=D_{-}$ and 
\be
K_{(2)}= 1,~~~~~U_{(2)}= M^2_{\rm eff}+\frac{{\rm k}^2}{a^2}\,. 
\ee
Here and in what follows  we denote, depending on the model,
\be                      \label{Meff}
{\rm model~I}:~~~M^2_{\rm eff}&=&M^2+\frac13\,\rho,~~~~\mh^2=M^2_{\rm eff},  \nonumber \\
{\rm model ~II}:~~~M^2_{\rm eff}&=&M^2-p,~~~~~~\mh^2=M^2+\rho. 
\ee
Here  $M_{\rm eff}$ is the effective mass  of the spin-2 particles,  while 
$\mh$ reduces to the Higuchi mass $M_{\rm H}$ in the Einstein  space limit, when $\rho=-p=\Lambda$. 
We notice that the effective mass depends on the background matter 
and in model I the spin-2 particles effectively become  heavier in regions of higher  background 
energy density $\rho$. A similar phenomenon is observed in the context of the massive 
bigravity theory \cite{Aoki:2017ffl}.  
In model II, curiously, $M^2_{\rm eff}$ decreases and  may become negative when $p$ grows. 

For the vector modes one has $Y={\rm V}_{+}$ or $Y={\rm V}_{-}$ and, defining  $\epsilon=\rho+p$, 
\be                               \label{Kv}
K_{(1)}=\frac{{\rm k}^2 m^4_{\rm H}}{m^4_{\rm H}+({\rm k}^2/a^2)(m_{\rm H}^2-\epsilon/2)},~~~~~~~~~
U_{(1)}=M^2_{\rm eff}\,{\rm k}^2. 
\ee

In the scalar sector one has $Y={\rm S}$ and the kinetic term 
\be                              \label{Ks}
K_{(0)}=\frac{3{\rm k}^4 m^4_{\rm H}(m_{\rm H}^2-2H^2)}
{(m_{\rm H}^2-2H^2)[9m_{\rm H}^4+6({\rm k}^2/a^2) (2m_{\rm H}^2-\epsilon )]
+4({\rm k}^4/a^4)(m_{\rm H}^2-\epsilon)
}.
\ee
The  potential $U_{(0)}$ in the scalar sector  is more complicated (see Eq.\eqref{C4} in  Appendix \ref{AppC}) 
but its asymptotic behavior is simple.
One has in all sectors 
\be
M^2_{\rm eff}~~\underset{{\rm k}\to 0}{\longleftarrow}~~~~ \frac{U}{K} ~~~~ 
\underset{{\rm k}\to\infty}{\longrightarrow}~~c^2\,\frac{{\rm k}^2}{a^2}
\ee
where $c$ is the sound speed. One finds for the tensors, 
vector and scalars, respectively, 
\be       \label{cs}
c_{(2)}^2&=& 1,   \nn  \\
c_{(1)}^2&=&\frac{M_{\rm eff}^2}{\mh^4}\,\left(\mh^2-\frac{\epsilon}{2}\right),   \\
c_{(0)}^2&=&\frac{(\mh^2-\epsilon)[\mh^4+(2H^2-4M^2_{\rm eff}-\epsilon)\mh^2+4H^2M^2_{\rm eff} ]   }
{3\mh^4(2H^2-\mh^2)}.  \nn
\ee
For the vectors and scalars one has 
$c^2<1$ but $c^2\to 1$  if $\rho\to 0$. 

For the Einstein space background one has 
$R_{\mu\nu}=\Lambda g_{\mu\nu}$  and $\rho=-p=\Lambda$ 
hence 
  $m^2_{\rm H}= M^2_{\rm H}$  and $2H^2= M^2_{\rm PM}=2\Lambda/3$. 
The above formulas  then imply that 
if $M_{\rm H}=0$ then $K_{(0)}=K_{(1)}=0$, therefore 
the scalar and  vector sectors become non-dynamical and only  the tensor modes propagate. 
The massless theory is recovered in this way. If $0<M_{\rm H}<M_{\rm PM}$ 
then $K_{(0)}<0$ (for ${\rm k}\to\infty$)  and the scalar polarization becomes 
a (Higuchi) ghost \cite{Higuchi}. If $M_{\rm H}=M_{\rm PM}$ then $K_{(0)}=0$
and  the scalar polarization 
 is non-dynamical (the PM case).

 All these features  are well known for massive gravitons in Einstein spaces. 
However,  for generic backgrounds, where $\rho,p$ are not constant, 
$m_{\rm H}$ and 
$H$ become functions of time, and it is not possible 
to have $K_{(1)}=0$ or $K_{(0)}=0$ for all time moments, whatever the value of the FP mass $M$ is. 
Therefore, neither the massless nor PM  cases are contained in the theory
for generic backgrounds -- the theory   always propagates  five polarizations. 
At most, there could be special backgrounds 
where spin-2 particles  become massless or PM for some values of $M$.

A direct inspection of Eqs.\eqref{Kv}--\eqref{cs} shows that 
if $\rho$ is small, 
$\rho\leq M^2$ 
 ($\bm{\rho}\leq M^2 M_{\rm Pl}^2$), then 
$K>0$ (for ${\rm k}\to\infty$) and $c^2>0$, 
hence the system is free of ghosts and tachyons. 
{The situation is more complex for large $\rho$. 
In model I  the kinetic term $K_{(0)}$  changes sign  for $\rho>3M^2$
 since $\mh^2<2H^2$ in this case, 
which corresponds to the Higuchi ghost. However, $c_{(0)}^2$ also 
changes sign at the same time (unless for $p/\rho=-1$)  so that  
the ghost and tachyon ``compensate each other", only changing 
the overall sign of the action. 
In model II  one always has  $\mh^2>2H^2$ and 
 the Higuchi ghost is absent, but since $M^2_{\rm eff}$ may be negative, 
 there could be tachyons  in the vector sector. 
 However, one finds in this case 
  that $K>0$ (always for ${\rm k}\to\infty$) and that $c^2>0$
for any $\rho$, provided that $p/\rho <-2/5$. 
Therefore, model II is  stable during the 
 inflationary stage, whereas model I is stable if the graviton mass is large enough, $M\geq H$. 
 Estimating that $\bm{\rho}\approx (10^{16}{\rm GeV})^4$ 
 at the beginning of the radiation-dominated stage \cite{weinberg}, 
 it follows that for $M\geq 10^{13}$ GeV one would have 
 ${\rho} \leq M^2$, and hence both models I and  II would  be stable at all times after the inflation.

A much milder bound $M\geq 10^{-3}$ eV is needed to insure that both models  are stable 
at present, that $\rho$ is small. 
Assuming that the $X_{\mu\nu}$ field couples only to the gravity 
and hence does not have other decay channels, 
it follows that it could be a part of the Dark Matter at present. 
 Massive spin-2 particles  as the DM candidates have actually been considered  before 
  \cite{Dubovsky:2004ud,Aoki:2016zgp,Babichev:2016bxi,Aoki:2017cnz},
 but only  our description  is consistent for arbitrary  backgrounds. 
 
 One should also say that the recent LIGO data  \cite{Abbott:2016blz} imply 
 that the graviton mass should be sufficiently small -- less than $10^{-22}$ eV
 \cite{deRham:2016nuf}. This seems to be in conflict with our estimate  $M\geq 10^{-3}$ eV. 
 However, the observational  bound 
applies rather to the mass of quanta of the background metric $g_{\mu\nu}$ and not to that of $X_{\mu\nu}$. 
As was mentioned above, it is consistent to consider  $X_{\mu\nu}$ as describing 
massive ``mesons" which may be not 
directly  interacting  with the ordinary matter and hence not seen by the  LIGO detector. 
Therefore  the bound does not apply to the FP mass $M$.

 It is also worth emphasising that, 
 since for the cosmological background 
 there are no  ghosts and tachyons,
  there is no superluminality problem in this case \cite{Deser:2012qx}.

  \section{Backreaction of massive spin-2 field}
  
  Apart from cosmology, the theory 
  of massive spin-2 field in curved space can have other applications. 
  For example, it can be used for 
  the holographic description of superconductors \cite{Benini:2010pr} 
  or electron-phonon interactions \cite{Baggioli:2014roa}.
  Up to now all applications have always been restricted to the Einstein spaces, 
  but in our theory this is no longer necessary. 
  
We have always assumed the background geometry to be fixed -- 
for example determined by Einstein equations with some matter source.   
At the same time, the massive spin-2 field can itself be the matter 
 source affecting the background. 
  To calculate its backreaction, one adds 
  the Einstein-Hilbert term to the action  \eqref{act} to obtain 
  \be                           \label{I}
  I=\frac12\int\left(M_{\rm Pl}^2\,R+X^{\nu\mu}E_{\mu\nu}\right)\sqrt{-g}\, d^4x.
  \ee
  Varying this with respect to the metric and $X_{\mu\nu}$ 
  yields the Einstein equations and the equations for $X_{\mu\nu}$ to be solved together, 
  \be                 \label{XE}
  M_{\rm Pl}^2\, G_{\mu\nu}=T_{\mu\nu}, ~~~~~
  E_{\mu\nu}=0,
  \ee 
  where the energy-momentum tensor 
  $T_{\mu\nu}$ is shown in  Appendix \ref{AppD}.  
   One should stress that, 
  irrespectively of whether it backreacts  or not, the $X_{\mu\nu}$ field always 
  propagates only 5 DoF.

 As the simplest application, we solved  equations \eqref{XE} in the homogeneous and isotropic sector, 
 with  $X_{\mu\nu}=X^{(0)}_{\mu\nu}$ given by \eqref{XX} for ${\rm k}=0$.
The goal was to see if the cosmology could be driven by the massive spin-2 field alone,  
as  happens in the  massive gravity models \cite{Volkov:2013roa}. 
However, our result was somewhat discouraging -- we found a solution only in model II and only for $M^2<0$: 
this is   the de Sitter space with $\Lambda=-3M^2>0$. 
  For this to be possible, one  should assume 
   $M^2$ to be negative, but such a theory would be very unstable since, for example, 
  $K$ and $c^2$ in \eqref{Kv}--\eqref{cs} would then be negative too. 
  We therefore conclude that the  theory  \eqref{I} cannot mimic  a  positive $\Lambda$-term.


  One can also study other solutions of equations   \eqref{XE}, 
  as for example black holes. 
 Interesting applications could  be found in connection with the phenomenon of superradiance 
of massive fields in the vicinity of spinning black holes \cite{Bardeen:1972fi,Starobinsky:1973aij} 
(see \cite{Brito:2015oca} for a recent review). 
The superradiance can 
lead to a spontaneous formation of massive clouds  evolving towards 
stationary {\it hairy} black holes \cite{Herdeiro:2014goa}.
 Such a ``spontaneous  bosonisation"  was actually predicted long ago \cite{Damour:1976kh},
 but only very recently the phenomenon 
 has been  confirmed by  numerical calculations \cite{East:2017ovw}. 
 For a spin-0 field the superradiance rate is not very high, but it  increases with spin  
\cite{Bardeen:1972fi,Starobinsky:1973aij}, 
which is why a  gravitating complex spin-1 field was considered in  \cite{East:2017ovw}. 
For a massive spin-2 field the superradiance should be faster still and 
one could expect hairy black holes to form spontaneously.

This suggests considering an extension of the theory \eqref{I} in which the field $X_{\mu\nu}$  is complex-valued, 
 \be                           \label{II}
  I=\frac12\int\left(M_{\rm Pl}^2\,R+\bar{X}^{\nu\mu}E_{\mu\nu} +   {X}^{\nu\mu}\bar{E}_{\mu\nu}   \right)\sqrt{-g}\, d^4x. 
  \ee
  Here the  bar denotes  complex conjugation and $E_{\mu\nu}\equiv \Delta_{\mu\nu}+{\cal M}_{\mu\nu}$
is expressed in terms of $X_{\mu\nu}$ by the same formulas as before. 
We expect this theory to admit stationary axially-symmetric black hole solutions 
supporting non-trivial massive hair of the form 
$
X_{\mu\nu}= e^{i\omega t+i m\varphi} \,{\bf X}_{\mu\nu}(r,\vartheta),
$   
so that there is a time-dependent spinning phase but the field amplitude is stationary. 
Such fields  with spinning phases are sometimes called massive clouds. 
Stationary black holes with scalar \cite{Herdeiro:2014goa} or vector  \cite{Herdeiro:2016tmi} 
clouds have been constructed 
explicitly. This suggests that the theory \eqref{II} could admit stationary black holes supporting tensor spin-2 massive clouds. 
One may expect such tensor clouds to form spontaneously due  to the superradiance  of  massive spin-2 particles.
This process could probably be simulated 
following the approach of  \cite{East:2017ovw}.

\section{Summary} 
We have constructed the exceptional theory of a free massive spin-2 field  in curved space. 
It is exceptional because it  propagates 5 DoF
for an arbitrary background geometry,  whereas almost all other known theories of this type 
propagate 5 DoF plus an additional ghostly  polarization. 
Only one other theory is exceptional in the same sense   -- it was recently 
constructed in  \cite{Bernard:2014bfa,Bernard:2015mkk,Bernard:2015uic}. That theory and our theory 
are probably 
equivalent since they are constructed in a similar way, but the equivalence is not manifest  since the
parameterizations of the two  theories are quite different.

Our  theory is described 
 by a non-symmetric tensor 
 $X_{\mu\nu}$ that 
fulfills equations $\Delta_{\mu\nu}+{\cal M}_{\mu\nu}=0$
where  the kinetic term $\Delta_{\mu\nu}$ and the mass term ${\cal M}_{\mu\nu}$ are defined by
\eqref{Del} and \eqref{Mb0}.  These equations 
imply 6 algebraic background-dependent 
conditions \eqref{sym} and five differential constraint \eqref{Cv1} and \eqref{C5}
 which together reduce the number of independent components of 
$X_{\mu\nu}$  from 16 to 5. This matches the number of polarisations of massive spin-2 particles. 

We emphasise once again that the property to propagate 5 (or less) DoF holds in our theory for 
{\it any} background geometry, whereas in the other known models this property 
holds only in  Einstein spaces. 

The mass term ${\cal M}_{\mu\nu}$ in our theory depends on four 
parameters $\beta_0,\beta_1,\beta_2,\beta_3$ and  on the background geometry 
via matrices $\gm_{\mu\nu}$, $\gam_{\mu\nu}$ algebraically 
 related to the background 
 by conditions \eqref{EG} or \eqref{EG1}. Resolving the latter with respect to $\gm_{\mu\nu}$, $\gam_{\mu\nu}$ 
 yields in general several solution branches and hence several different mass terms ${\cal M}_{\mu\nu}$. 
 In all cases ${\cal M}_{\mu\nu}$ is a linear combination of the background metric and of 
 powers of the background Ricci tensor, as shown by  Eq.\eqref{mass}. 
 
 Different choices of the  mass term correspond to different theories. 
 All these theories propagate 5 DoF  but their other properties are not necessarily the same. 
 For two special theories, that we call models I and II, the mass term is a linear function of the 
 background Ricci tensor (Eqs.\eqref{2},\eqref{2a}). The only free parameter left in this  case is the FP mass $M$. 
 Within these two models, we constructed the 
 general solution for $X_{\mu\nu}$ on a homogeneous and isotropic cosmological background and found 
  this solution to be stable. 
 Therefore,  massive spin-2 particles could potentially contribute to the Dark Matter.

Summarizing, we presented the exceptional  theory of a massive spin-2 field 
in curved space parameterized in an unusual way 
--  in terms of a non-symmetric rank-2 tensor. Our main goal 
was to show that the theory is self-consistent and that the number of independent DoF is indeed 5. 
We have shown this by counting the constraints and also by counting the independent modes in the 
general solution.

  \section*{Acknowledgements}    
 We thank Arkady Tseytlin and Matteo Beccaria
 for  critical remarks and confirming our formula \eqref{act} for the action, and also C\'edric Deffayet
 for discussions. 
 We also thank 
Eugen Radu and Carlos Herdeiro  for explaining to us the 
 modern aspects of the 
 superradiance phenomenon. 
 M.S.V. was partly supported by the Russian Government Program of Competitive Growth 
of the Kazan Federal University.

\appendix

\setcounter{section}{0}
\setcounter{equation}{0}
\setcounter{subsection}{0}

\section{Solution for $\gamma_{\mu\nu}$ \label{App0}}

\renewcommand{\theequation}{A.\arabic{equation}} 

\renewcommand{\theequation}{A.\arabic{equation}} 

Here we illustrate how the background equations \eqref{EG1} are solved in the simple 
 case where $\beta_3=0$. 
For $\beta_3\neq 0$ the procedure is similar but the formulas are more complicated. 
Introducing  matrices $\hat{\gamma}=\gamma^\mu_{~\nu}$ and $\hat{R}=R^\mu_{~\nu}$
we denote by $\e_k\equiv \e_k(\hat{R})$ the 
scalar invariants defined by \eqref{inv}. 
Equations \eqref{EG1} with $\beta_3=0$ can be written as 
\be                \label{eqM}
\hat{R}-\frac12\, \e_1+\beta_0+\beta_1\Big( [\hat{\gamma}] -\hat{\gamma}  \Big )
+\beta_2 \Big(
\hat{\gamma}^2-[\hat{\gamma}]\hat{\gamma} +\frac12([\hat{\gamma}]^2-[\hat{\gamma}^2]
\Big )=0.
\ee
Viewed as algebraic equations for $\hat{\gamma}$, their solution has to be of  the form 
\be             \label{solg}
\hat{\gamma}=y_0+y_1\,\hat{R}+y_2\,\hat{R}^2+y_3\,\hat{R}^3 
\ee
so that 
$
[\hat{\gamma}]=4y_0+y_1\,[\hat{R}]+y_2\,[\hat{R}^2]+y_3\,[\hat{R}^3]
$
where 
\be
[\hat{R}]=\e_1,~~~[\hat{R}^2]=\e_1^2-2\e_2,~~~~[\hat{R}^3]=\e_1^3-3\e_1\e_2+3\e_3\,.
\ee
The next step is to eliminate the higher powers of $\hat{R}$ from 
$\hat{\gamma}^2=(y_0+y_1\,\hat{R}+y_2\,\hat{R}^2+y_3\,\hat{R}^3)^2$ 
by using  the Hamilton-Cayley relation 
\be
\hat{R}^4=\e_1\hat{R}^3-\e_2\hat{R}^2+\e_3\hat{R}-\e_4. 
\ee
This yields $\hat{\gamma}^2=A_0+A_1\hat{R}+A_2\hat{R}^2+A_3\hat{R}^3$ with 
\be
A_0&=&
-y_3^2 \e_1^2 \e_4-2y_2y_3\e_1\e_4+y_3^2\e_2\e_4-2y_1y_3\e_4-y_2^2\e_4+y_0^2\,,  \nn  \\
A_1&=&
y_3^2 \e_3\e_1^2+(2y_2y_3\e_3-y_3^2\e_4)\e_1-y_3^2 \e_2\e_3+(2y_1y_3+y_2^2)\e_3-2y_2y_3\e_4+2y_0y_1\,,
 \nn  \\
A_2&=&
(y_3^2\e_3-2y_2y_3\e_2)\e_1-y_3^2 \e_1^2\e_2+y_3^2\e_2^2-(2y_1y_3+y_2^2)\e_2  \nn \\
&&+2y_2y_3\e_3-y_3^2\e_4+2y_0y_2+y_1^2\,,
     \\
A_3&=&
y_3^2\e_1^3+2y_2 y_3 \e_1^2-2y_3^2 \e_1\e_2+(y_2^2+2y_1y_3)\e_1-2y_2y_3\e_2+y_3^2\e_3
+2y_0y_3+2y_1y_2 \,,    \nn
\ee
while $[\hat{\gamma}^2]=4A_0+A_1[\hat{R}]+A_2[\hat{R}^2]+A_3[\hat{R}^3]$.  
One can similarly 
express $\hat{\gamma}^3$, but for $\beta_3=0$ this is not necessary. 

Inserting 
$\hat{\gamma}$, $\hat{\gamma}^2$, $[\hat{\gamma}]$, $[\hat{\gamma}^2]$ to \eqref{eqM}
and setting to zero the coefficients in front of the matrices $\hat{1}=\hat{R}^0$, $\hat{R}$, $\hat{R}^2$, 
$\hat{R}^3$ yields four algebraic relations 
\be
\Big (
2y_1y_2-2y_0y_3+\e_1\,(y_1y_3-y_2^2)+\e_1^2\,y_2y_3+(\e_1\e_2-2\e_3)y_3^2
\Big )\beta_2-\beta_1 y_3=0, \nn  \\
\nn \\
\Big(
y_1^2-\e_1\,y_1y_2-2\e_2\,y_1y_3+(\e_2-\e_1^2)y_2^2+(\e_1\e_2-\e_3-\e_1^3)y_2y_3-2y_0y_2& &  \nn  \\
+(\e_2^2+\e_1\e_3-\e_2\e_1^2-\e_4)y_3^2
\Big)\beta_2-\beta_1 y_2=0,  \nn  \\ 
\nn  \\
\Big (
(2\e_2-\e_1^2)y_1 y_2 -\e_1 y_1^2 +(3\e_1 \e_2 -\e_1^3-\e_3 )y_1 y_3 -2 y_0y_1\nn \\
+\e_3 y_2^2+2(\e_1\e_3-\e_4)y_2y_3 
+(\e_1^2 \e_3-\e_1\e_4-\e_2\e_3)y_3^2
\Big )\beta_2-\beta_1 y_1+1=0,  \nn \\
 \nn  \\
\Big (
3y_0^2+2\e_1 y_0y_1+2(\e_1^2-2\e_2)y_0y_2+2(\e_1^3-3\e_1\e_2+3\e_3)y_0y_3 
+\e_2 y_1^2  \nn   \\
+(\e_1\e_2-3\e_3)y_1y_2  
+(\e_1^2\e_2-\e_1\e_3-2\e_2^2+2\e_4)y_1y_3 
+(\e_4+\e_2^2-2\e_1\e_3)y_2^2   \nn   \\
+(\e_1\e_2^2-2\e_1^2\e_3+3\e_1\e_4-\e_2\e_3)y_2y_3  \nn \\
+(3\e_3^2-2\e_2\e_4+\e_2^3-3\e_1\e_2\e_3+2\e_1^2\e_4)y_3^2
\Big )\beta_2   \nn   \\
+ \Big (
(3\e_3-3\e_1\e_2+\e_1^3)y_3+(\e_1^2-2\e_2)y_2+\e_1y_1+3y_0
\Big )\beta_1+\beta_0-\frac12\,\e_1=0.\nn \\
\ee
These determine the four coefficients $y_0$, $y_1$, $y_2$, $y_3$
in the solution \eqref{solg}.


\setcounter{equation}{0}

\section{Scalar constraint \label{AppA}}

\renewcommand{\theequation}{B.\arabic{equation}}

Here is the off-shell value of the scalar constraint defined by Eq.\eqref{C5},
\be
\mathcal{C}_{5} &=&
\mathfrak{A}^{\lambda \sigma \alpha \beta} \nabla_{\lambda} \nabla_{\sigma} X_{\alpha \beta} 
+ \mathfrak{B}^{\sigma \alpha \beta} \nabla_ {\sigma} X_{\alpha \beta} 
+ \mathfrak{C}^{\alpha \beta} X_{\alpha \beta}\,,
\ee
where  the coefficients are 
\be   
\mathfrak{A}^{\lambda \sigma \alpha \beta} &=&\beta_{3} |\gamma|\Sigma^{\lambda \sigma \alpha \beta} 
+ \beta_{3} \frac{|\gamma|}{g^{00}} \Sigma^{00 \mu \nu} \Big [ \frac{1}{2} g_{\mu \nu} 
\big ( g^{\sigma \lambda} g^{\alpha \beta} - g^{\alpha \lambda} g^{\beta \sigma} \big )  \nn  \\
&&+ \frac{g_{\mu  \nu}}{2 \, g^{00}} \big ( g^{\lambda \beta} g^{0 \sigma} g^{0 \alpha}  
    - g^{\sigma \lambda} g^{0 \alpha} g^{0 \beta}   
 + g^{\alpha \lambda} g^{0 \sigma} g^{0 \beta} 
  - g^{\alpha \beta} g^{0 \sigma} g^{0 \lambda} \big ) \nn \\
&&  +  g^{\alpha \lambda} \delta^{\sigma}_{(\mu} \delta^{\beta}_{\nu)} 
  + g^{\beta \lambda} \delta^{\sigma}_{(\mu} \delta^{\alpha}_{\nu)} 
  - g^{\sigma \lambda} \delta^{\alpha}_{(\mu} \delta^{\beta}_{\nu)}
   -  g^{\alpha \beta} \delta^{\lambda}_{(\mu} \delta^{\sigma}_{\nu)} \Big ],   \nn \\
   \nn  \\
 \mathfrak{B}^{\sigma \alpha \beta}  &=&  \beta_{1} \big [ 
 \Gamma^{\sigma \lambda} ( \nabla^{\alpha} \gamma^{\beta}_{\, \lambda} 
 - \nabla_{\lambda} \gamma^{\alpha \beta} ) 
 + \Gamma^{\sigma \beta} ( \nabla^{\lambda} \gamma^{\alpha}_{\, \lambda} 
 - \nabla^{\alpha} [\gamma] ) \big ]   \nn  \\
 &+& \beta_{2} \Big [g^{\alpha \beta} ( \nabla_{\lambda} \gamma^{\lambda \sigma} -  \nabla^{\sigma} [\gamma])
  - g^{\alpha \sigma} ( \nabla_{\lambda} \gamma^{\lambda \beta}   
 -  \nabla^{\beta} [\gamma] )   \nn \\
 &&~~~+ \nabla^{\sigma} \gamma^{\alpha \beta} 
 -  \nabla^{\beta} \gamma^{\alpha \sigma} 
 + \Gamma^{\sigma \nu} \big ( \nabla^{\mu} H^{\alpha \beta}_{\, \mu \nu} 
 - \nabla^{\alpha} \{Q^{\beta}_{\, \nu} 
 - \frac{1}{2} [Q] \delta^{\beta}_{\nu} \} \big ) \Big ] \nn \\
& + &\beta_{3} \big [ \nabla_{\lambda} ( |\gamma|\Sigma^{\lambda \sigma \alpha \beta} ) - g^{\alpha \beta} \Gamma^{\sigma \lambda} \nabla^{\rho} ( |\gamma|\Gamma_{\rho \lambda} ) \big ],    \nn   \\
\nn  \\
\mathfrak{C}^{\alpha \beta}
&=&  \beta_{1} \Big [ R^{\alpha \beta} + \nabla_{\sigma} \Big\{ \Gamma^{\sigma \lambda} ( \nabla^{\alpha} \gamma^{\beta}_{\, \lambda} - \nabla_{\lambda} \gamma^{\alpha \beta} ) + \Gamma^{\sigma \beta} ( \nabla^{\lambda} \gamma^{\alpha}_{\lambda} - \nabla^{\alpha} [\gamma] ) \Big\} \Big ]    \nn   \\
&+& \beta_{2} \Big [ [\gamma] R^{\alpha \beta} - 2 \, R^{\alpha \lambda} \gamma^{\beta}_{\, \lambda}  
- R^{\beta \lambda} \gamma^{\alpha}_{\, \lambda} + R^{\beta \lambda \alpha \sigma} \gamma_{\lambda \sigma}  \nn \\
&& ~~~ +\nabla_{\lambda} \big ( \Gamma^{\lambda \nu} \big \{ \nabla^{\mu} H^{\alpha \beta}_{\, \mu \nu} - \nabla^{\alpha}(Q^{\beta}_{\, \nu} - \frac{1}{2} [Q] \delta^{\beta}_{\nu}) \big \} \big ) \big ] \nn \\
&+& \beta_{3} \Big [ - g^{\alpha \beta} \nabla_{\lambda} \{ \Gamma^{\lambda \sigma} \nabla^{\rho} ( |\gamma|\Gamma_{\rho \sigma} ) \}   \nn  \\
&&~~~+ |\gamma|\frac{\Sigma^{00 \mu \nu}}{g^{00}} \Big ( \frac{1}{2} g_{\mu \nu} \{ R^{\alpha \beta} - \frac{2}{g^{00}} g^{0 \beta} R^{\alpha 0} \} - 2 \, R^{\alpha}_{\, (\mu} \delta^{\beta}_{\nu)} \Big ) \Big ]\nn \\
&-& \frac{3}{2}\, \beta^2_{1} \gamma^{\alpha \beta} + 2 \, \beta_{1} \beta_{2} Q^{\alpha \beta}   \nn  \\
&+& \beta_{1} \beta_{3} \, |\gamma| \Big [ \frac{1}{2} ( \Gamma^{\alpha \beta} - g^{\alpha \beta} [\Gamma] ) + \frac{\Sigma^{00 \mu \nu}}{g^{00}} \big ( \gamma^{\alpha}_{\mu} \delta^{\beta}_{\nu} + \frac{g_{\mu \nu}}{2 \, g^{00}} \gamma^{\alpha 0} g^{\beta 0} \big ) \Big ]  \nn \\
& + &2 \, \beta^2_{2} |\gamma|\Big [  \Gamma^{\alpha \beta} - g^{\alpha \beta} [\Gamma]  \Big ]   \nn  \\
&+& \beta_{2} \beta_{3} |\gamma|\Big [ - 3 \, g^{\alpha \beta} 
+ \frac{\Sigma^{00 \mu \nu}}{g^{00}} \big ( H^{\alpha \beta}_{\mu \nu} 
+ \frac{g_{\mu \nu}}{2 \, g^{00}} H^{00 \alpha \beta} - g_{\mu \nu} Q^{\alpha \beta} \big ) \Big ] \nn \\
& + &\beta^2_{3} |\gamma|^2 \frac{\Sigma^{00 \mu \nu}}{g^{00}} \Big [ \delta^{\alpha}_{\mu} \Gamma^{\beta}_{\nu} - g^{\alpha \beta} \Gamma_{\mu \nu} + \frac{1}{2} g_{\mu \nu} \Gamma_{\sigma \lambda} (g^{\alpha \beta} \hh^{\lambda \sigma} - g^{\beta \sigma} \hh^{\alpha \lambda} ) \Big ]. 
\ee
Here, as usual, $[M]$ denotes the trace, 
the tensors $Q^{\beta}_{\alpha}$ and $H^{\alpha \beta}_{\mu \nu}$ are defined in \eqref{QQ} and \eqref{HH}, 
and we have introduced 
\begin{equation}
\hh^{\mu\nu}=g^{\mu\nu}-\frac{1}{g^{00}}\,g^{0\mu}g^{0\nu}\,,~~~~~
\Sigma^{\mu \nu \alpha \beta} = \Gamma^{\mu \beta} \Gamma^{\nu \alpha} - \Gamma^{\mu \nu} \Gamma^{\alpha \beta}.
\end{equation}

\setcounter{equation}{0}

\renewcommand{\theequation}{C.\arabic{equation}} 

\section{ Constraints  in models I and II \label{AppB}}

Here we show the derivation of the 
constraints in models I and II expressed  by Eqs.\eqref{vI}, \eqref{sI}, \eqref{vII}, \eqref{sII}  in the main text. 
Using 
$$(\nabla_\mu\nabla_\nu-\nabla_\nu\nabla_\mu)X^\alpha_{~\beta}
=R^\sigma_{~\beta\nu\mu}X^\alpha_{~\sigma}
-R^\alpha_{~\sigma\nu\mu}X^\sigma_{~\beta}\,,
$$
a direct calculation yields the following result for the divergence 
of $\Delta_{\mu\nu}$ defined by Eq.\eqref{Del} in the main text:
\be              \label{20aa}
\nabla^\mu\Delta_{\mu\nu}&=&
\bm{\gamma}_{\nu\beta}(\nabla_\alpha X^{\alpha\beta}-\nabla^\beta X)\, \nn \\
&+&\bm{\gamma}_{\alpha\beta}(\nabla_\nu X^{\alpha\beta}-\nabla^\alpha X^\beta_{~\nu})  \nn \\
&+&X^{\alpha\beta}\nabla_\alpha G_{\beta\nu}
\ee
with 
\be 
\bm{\gamma}_{\mu\nu}\equiv R_{\mu\nu}+\bm{\phi}\, g_{\mu\nu}\,,
\ee
where $\bm{\phi}$ 
can be set to any value  because the part of $\bm{\gamma}_{\mu\nu}$ proportional to $g_{\mu\nu}$ 
cancels in \eqref{20aa}.  In particular, one can adjust $\bm{\phi}$ such that 
the tensor $\bm{\gamma}_{\mu\nu}$ will correspond either to 
that given by Eq.\eqref{gamI} in model I or to that expressed by \eqref{gamII} in model II. 
If $R_{\mu\nu}=\Lambda g_{\mu\nu}$ then $\bm{\gamma}_{\mu\nu}\sim g_{\mu\nu}$ 
and \eqref{20} yields $\nabla^\mu\Delta_{\mu\nu}=0$. 

The divergence of 
${\cal M}_{\mu\nu}= \gmm_{\mu\alpha} X^\alpha_{~\nu}
-g_{\mu\nu}\,\gmm_{\alpha\beta}X^{\alpha\beta}$ 
in model I  (see Eq.\ref{2}) is 
\be              \label{21} 
\nabla^\mu {\cal M}_{\mu\nu}=\bm{\gamma}_{\alpha\beta}(\nabla^\alpha X^\beta_{~~\nu}-\nabla_\nu X^{\alpha\beta})  \nn \\
+X^\alpha_{~~\nu}\nabla^\mu \bm{\gamma}_{\mu\alpha}-X^{\alpha\beta}\nabla_\nu \bm{\gamma}_{\alpha\beta}\,.
\ee
Adding this up with \eqref{20aa},
the second line on the right in \eqref{20aa} cancels against the first line in \eqref{21}, yielding 
\be
{\cal C}_\nu &\equiv& \nabla^\mu E_{\mu\nu}\equiv 
\nabla^\mu(\Delta_{\mu\nu}+{\cal M}_{\mu\nu})  \nn  \\
&=&\bm{\gamma}_{\nu\beta}(\nabla_\alpha X^{\alpha\beta}-\nabla^\beta X) 
+X^{\alpha\beta}(\nabla_\alpha G_{\beta\nu}-\nabla_\nu \bm{\gamma}_{\alpha\beta})
+X^\alpha_{~~\nu}\nabla^\mu \bm{\gamma}_{\mu\alpha}  \, ,
\ee
which reproduces Eq.\eqref{vI} in the main text. 
Multiplying this by the inverse of $(\bm{\gamma}^{-1})^{\rho\nu}$,
acting with $\nabla_\rho$ and combining with the trace $E^\mu_{~\mu}$ reproduces 
Eq.\eqref{sI} in the main text. 

The divergence of ${\cal M}_{\mu\nu}= -X_\mu^{~\alpha} \gmm_{\alpha\nu} 
+X\gmm_{\mu\nu}$ in model II is 
\be              \label{22} 
\nabla^\mu {\cal M}_{\mu\nu}=\bm{\gamma}_{\nu\beta}(\nabla^\beta X-\nabla_\alpha X^{\alpha\beta}) \nn \\
+X\nabla^\mu\bm{\gamma}_{\mu\nu}-X^{\alpha\beta}\nabla_\alpha\bm{\gamma}_{\beta\nu}\,,
\ee
where $\bm{\gamma}_{\mu\nu}=G_{\mu\nu}-M^2g_{\mu\nu}$. 
Adding this up with \eqref{20aa}, the first and third 
lines on the right in \eqref{20aa} cancel against \eqref{22}, hence 
\be              \label{20a}
{\cal C}_\nu\equiv 
\nabla^\mu(\Delta_{\mu\nu}+{\cal M}_{\mu\nu})=
\bm{\gamma}_{\alpha\beta}(\nabla_\nu X^{\alpha\beta}-\nabla^\alpha X^\beta_{~\nu}), 
\ee
which reproduces Eq.\eqref{vII} in the main text. 
Multiplying this by $\bm{\gamma}^{\rho\nu}$  yields 
\be
\bm{\gamma}^{\rho\nu}{\cal C}_\nu&=& \bm{\gamma}^{\rho\nu}\bm{\gamma}_{\alpha\beta}(\nabla_\nu X^{\alpha\beta}-\nabla^\alpha X^\beta_{~\nu}) 
=(\bm{\gamma}^{\rho\nu}\bm{\gamma}^{\alpha\beta}-\bm{\gamma}^{\rho\beta}\bm{\gamma}^{\nu\alpha})\nabla_\nu X_{\alpha\beta}\,. 
\ee
Acting on this with $\nabla_\rho$ one obtains 
\be                \label{BB7}
\nabla_\rho(\bm{\gamma}^{\rho\nu}{\cal C}_\nu)&=&(\bm{\gamma}^{\rho\nu}\bm{\gamma}^{\alpha\beta}-\bm{\gamma}^{\rho\beta}\bm{\gamma}^{\nu\alpha})
\nabla_\rho\nabla_\nu X_{\alpha\beta}  
+\nabla_\rho(\bm{\gamma}^{\rho\nu}\bm{\gamma}^{\alpha\beta}-\bm{\gamma}^{\rho\beta}\bm{\gamma}^{\nu\alpha})
\nabla_\nu X_{\alpha\beta} \nn \\
&=&(\bm{\gamma}^{00}\bm{\gamma}^{\alpha\beta}-\bm{\gamma}^{0\beta}\bm{\gamma}^{0\alpha})\ddot{X}_{\alpha\beta}+\ldots \\ \nn
&=&(\bm{\gamma}^{00}\bm{\gamma}^{ik}-\bm{\gamma}^{0i}\bm{\gamma}^{0k})\ddot{X}_{ik}+\ldots 
\ee
where 
the dots denote   terms not containing second time derivatives of $X_{\alpha\beta}$. One can now 
repeat  the general arguments given between Eq.\eqref{TTT} and Eq.\eqref{b33} in the main text 
to obtain 
\be
\ddot{X}_{(ik)}
 &=&\frac{1}{g^{00}}\left(\frac{1}{2}\,{g_{ik}}\,\hh^{nm} E_{nm}-E_{ik}  \right)+\ldots
\ee
with $\hh^{nm}=g^{nm}-g^{0n}g^{0m}/g^{00}$ and to conclude that 
the second time derivatives in \eqref{BB7} are exactly the same as in 
\be
\frac{1}{g^{00}}(\bm{\gamma}^{00}\bm{\gamma}^{\alpha\beta}
-\bm{\gamma}^{0\alpha}\bm{\gamma}^{0\beta})\left(\frac{1}{2}\,{g_{\alpha\beta}}\,
\left(E^\alpha_{~\alpha}-\frac{1}{g^{00}}\, E^{00} \right)
-E_{\alpha\beta}  \right).       \label{BB8}
\ee
Therefore, the difference of \eqref{BB7} abd \eqref{BB8} does not contain second time derivatives, which 
 yields Eq.\eqref{sII}  in the main text.

\section{Solution in the expanding universe \label{AppC}}

\setcounter{equation}{0}
\renewcommand{\theequation}{D.\arabic{equation}}

Inserting  the cosmological metric \eqref{cosm} and the harmonic decomposition \eqref{XX0}--\eqref{XX}
for $X_{\mu\nu}$ to the equations $E_{\mu\nu}\equiv \Delta_{\mu\nu}+{\cal M}_{\mu\nu}$ 
with $\Delta_{\mu\nu}$ given by \eqref{Del} and ${\cal M}_{\mu\nu}$ defined 
either  by \eqref{2} or by \eqref{2a}, the equations split into three independent sectors. 

The tensor sector contains only two amplitudes ${\rm D}_{+}$ and ${\rm D}_{-}$ whose equations 
can be obtained  by varying the effective action 
\eqref{act1} in the main text.

The vector sector contains 6 amplitudes, 4 of which, $W^\pm_{\pm}$, can be 
expressed by virtue of the field equations in terms of two independent ${\rm V}_{+}$ and ${\rm V}_{-}$  as
\be
W^{+}_\pm =\frac{{\rm P}^2 m_{\rm H}^2\, \dot{\rm V}_\pm}{
m_{\rm H}^4+{\rm P}^2(m_{\rm H}^2-\epsilon/2)
},   ~~~~
W^{-}_\pm =\frac{{\rm P}^2\,[m_{\rm H}^2-\epsilon]\, \dot{\rm V}_\pm}{
m_{\rm H}^4+{\rm P}^2(m_{\rm H}^2-\epsilon/2)
}.   
\ee 
Here  $m_{\rm H}$ is defined in Eq.\eqref{Meff}  in the main text, $\epsilon=\rho+p$,  and ${\rm P}={\rm k}/a$ 
is the physical momentum.  The equations 
for ${\rm V}_\pm$ reduce to those obtained by varying the effective action 
\eqref{act1} in the main text.

Finally, the field equations imply that 
the four scalar amplitudes $S^\pm_\pm$ in \eqref{XX} can be expressed in terms of one single 
${\rm S}$ by the following relations: 
\be
S^{-}_{+}&=&\frac{m_{\rm H}^2-\epsilon}{m_{\rm H}^2}\, S^{+}_{-}\,, \nn  \\
S^{+}_{-}&=&\frac{2}{m_{\rm H}^2}\,\left(\dot{S}^{-}_{-}+a^2 H S^{+}_{+}\right)    ,  \\
S^{+}_{+}&=&-\frac{1}{Ha^2}\,\dot{S}^{-}_{-}     
+\frac{2Hm_{\rm H}^4 {\rm P}^2\,\dot{\rm S}+m_{\rm H}^6 {\rm P}^2\, {\rm S}-m_{\rm H}^4 (2{\rm P}^2+3 m_{\rm H}^2) S^{-}_{-}/a^2   }{
2H^2[3m_{\rm H}^4+2{\rm P}^2(2m_{\rm H}^2-\epsilon)]},  \nn \\
S^{-}_{-}&=&a^2{\rm P}^2\frac{  \left\{
2{\rm P}^2[(\mh^2-2H^2)(2\mh^2-\epsilon)-\mh^4]+3\mh^4(\mh^2-2H^2)
\right\}{\rm S}     
 -4H\mh^2 {\rm P}^2\,\dot{\rm S }
}{4{\rm P}^4(\mh^2-\epsilon)+6{\rm P}^2(2\mh^2-\epsilon)(\mh^2-2H^2)+9\mh^4(\mh^2-2H^2)}. \nn 
\ee
It is crucial that all four  $S^\pm_\pm$ 
are expressed  in terms of one single ${\rm S}$ that  fulfills the master 
equation obtainable  by varying the effective action \eqref{act1} in the main text.
This shows that there is only one dynamical  DoF in the scalar sector. 
Therefore,  
together with the tensor and vector modes, the theory propagates 5 DoF. 

The kinetic term $K_{(0)}$ for the scalars is given by Eq.\eqref{Ks} 
while the potential term is 
\be                    \label{C4}
U_{(0)}=\frac{b_0+b_2 {\rm P}^2+b_4 {\rm P}^4+b_6 {\rm P}^6}{C\, (c_0+c_2{\rm P}^2+c_4{\rm P}^4) }\, K_{(0)}
\ee
where 
\be
C&=&3\mh^4(\mh^2-2H^2),  ~~~~~~
c_0=9\mh^4(\mh^2-2H^2),  \nonumber \\
c_2&=&6(\mh^2-2H^2)(2\mh^2-\epsilon),  ~~~~~~
c_2=4(\mh^2-\epsilon), 
\ee
and 
\be
b_0&=&27\,\mh^8 M^2_{\rm eff}(\mh^2-2H^2)^2,   \\
b_2&=&9\,\mh^4(\mh^2-2H^2)^2[4M^2_{\rm eff}(2\mh^2-\epsilon)-\mh^4  ],   \nonumber   \\
b_4&=&6\,\mh^4\left[8\mh^6-(20H^2+9\epsilon)\mh^4\right. \nonumber  \\
&+&\left.(8H^4+20H^2\epsilon+2H\dot{p}+\epsilon^2)\mh^2-4H^2(H\dot{p}+\epsilon^2)\right]       \nonumber \\
&+&12(M^2_{\rm eff}-\mh^2)\left[5\mh^8-6(2H^2+\epsilon)\mh^6\right.  \nonumber \\
&+& (8H^4+14H^2\epsilon+\epsilon^2)\mh^4 \nonumber \\
&-&\left.4H^2\epsilon(2H^2+\epsilon)\mh^2+4H^4\epsilon^2\right],  \nonumber \\
b_6&=&4(\mh^2-\epsilon)^2[4M^2_{\rm eff}(\mh^2-H^2)+\mh^2(\epsilon-2H^2-\mh^2)].  \nonumber 
\ee
Notice that these expressions contain $\dot{p}$ and hence the third derivative of the background scale factor $a(t)$. 
The ratio $c^2=b_6/(Cc_4)$ is the sound speed  expressed by Eq.\eqref{cs} in the main text.

\section{Energy-momentum tensor of massive spin-2 field \label{AppD}}

\setcounter{equation}{0}

\renewcommand{\theequation}{E.\arabic{equation}} 
Varying  the action \eqref{act} 
with respect to the spacetime metric, 
\be
\delta I=-\frac12\int T_{\mu\nu}\,\delta g^{\mu\nu}\,\sqrt{-g}\, d^4x, 
\ee
determines the energy-momentum tensor. 
It has a somewhat complicated structure, 
  partly  due to the non-minimal terms like $X^{\mu\nu}R^\sigma_{\nu}X_{\sigma\mu}$ 
  in the action. 
A straightforward (but lengthy)  calculation yields 
  in model I
 \be
T_{\mu \nu} &=& \nabla_{\lambda} \h_{\alpha \beta} \nabla_{\sigma} \h_{\rho \tau} \mathcal{A}^{\, \, \, \; \; \lambda \alpha \beta \sigma \rho \tau}_{\mu \nu} 
+  \h_{\alpha \beta} \nabla_{\lambda} \nabla_{\sigma} \h_{\rho \tau} \mathcal{B}^{\, \, \, \; \; \lambda \alpha \beta \sigma \rho \tau}_{\mu \nu} + R_{\alpha \beta} X^{\alpha}_{\, \, \, \, \mu} X^{\beta}_{\, \, \, \, \nu}   \nn \\
&+& 2\, R_{\alpha (\mu} X^{\, \, \, \, \beta}_{\nu)} X^{\alpha}_{\, \, \, \; \beta} 
 -\frac{1}{2} g_{\mu \nu} R_{\alpha \beta} X^{\alpha \lambda} X^{\beta}_{\, \, \, \, \lambda}  
 +\frac{1}{2} {\DAlambert} ( X_{\mu \lambda} X^{\, \, \, \, \lambda}_{\nu} )   \nn  \\
&+& \frac{1}{2} g_{\mu \nu} \nabla_{\alpha} \nabla_{\beta} (X^{\alpha \lambda} X^{\beta}_{\, \, \, \, \lambda}) - \nabla_{\alpha} \nabla_{(\mu}( X_{\nu) \beta} X^{\alpha \beta}) 
+ \frac{1}{6} \Big [  R_{\mu \nu} (X^{\alpha \beta} X_{\beta \alpha} - [X]^2)  \nn  \\
&-& \nabla_{(\mu} \nabla_{\nu)} (X^{\alpha \beta} X_{\beta \alpha} - [X]^2) + g_{\mu \nu} 
{\DAlambert} (X^{\alpha \beta} X_{\beta \alpha} - [X]^2) \Big ]  \\
& -& 2 \, \Big (M^2 - \frac{R}{6}\Big ) [ g_{\lambda\sigma}X^{~~\lambda}_{(\mu} X^\sigma_{~~\nu)} - [X] X_{(\mu \nu)} - \frac{1}{4} g_{\mu \nu} (X^{\alpha \beta} X_{\beta \alpha} - [X]^2) ],    \nn
\ee
and in model II
\be
T_{\mu \nu} &=& 
\nabla_{\lambda} \h_{\alpha \beta} \nabla_{\sigma} \h_{\rho \tau} \mathcal{A}^{\, \, \, \; \; \lambda \alpha \beta \sigma \rho \tau}_{\mu \nu} 
+  \h_{\alpha \beta} \nabla_{\lambda} \nabla_{\sigma} \h_{\rho \tau} \mathcal{B}^{\, \, \, \; \; \lambda \alpha \beta \sigma \rho \tau}_{\mu \nu} 
+ R_{\alpha \beta} \h^{\alpha}_{\, \, \, \, ( \mu} X^{\beta}_{\, \, \, \, \nu)}  \nn \\
&& + R_{\alpha \beta} X^{\, \, \, \,  \alpha}_{( \mu} X^{\beta}_{\, \, \, \, \nu)}  
+  R_{\beta (\mu} \h_{\nu) \alpha} X^{\alpha \beta} 
 + R_{\beta (\mu} X_{\nu) \alpha} \h^{\alpha \beta}   \nn  \\
&& + R_{\beta (\mu} ( X_{\alpha \nu )} X^{\beta \alpha} 
 + X_{\nu \alpha)} X^{\alpha \beta} )   
-\frac{1}{2} g_{\mu \nu} R_{\alpha \beta} X^{\alpha \lambda} X^{\, \, \, \, \beta}_{\lambda}   \nn \\
&& -\frac{1}{2} g_{\mu \nu} R_{\alpha \beta} X^{\alpha \lambda} \h^{\beta}_{\, \, \, \, \lambda} 
 + g_{\mu \nu} R_{\alpha \beta} X^{\alpha \beta} [X] 
- 2 R_{\lambda (\mu} \h^{\, \, \, \, \lambda}_{\nu)}[X] 
- R_{\alpha \beta} X^{\alpha \beta} \h_{\mu \nu}  \nn \\
& &+ \frac{1}{2} \,
{\DAlambert} [ X_{(\mu \lambda} \h^{\, \, \, \, \lambda}_{\nu)} 
+ X_{\lambda (\mu} X^{\, \, \, \, \lambda}_{\nu)} 
- [X] \h_{\mu \nu}]   \nn  \\
&&   + \frac{1}{2} g_{\mu \nu} \nabla_{\alpha} \nabla_{\beta} [X^{\alpha \lambda} \h^{\beta}_{\, \, \, \, \lambda} 
+ X^{\lambda \alpha} X^{\beta}_{\, \, \, \, \lambda} 
- 2 [X] X^{\alpha \beta}]       \\
&&- \frac{1}{2} \nabla_{\alpha} \nabla_{(\mu}[ X_{\nu) \beta} \h^{\alpha \beta} 
+ X_{\beta \nu)} X^{\alpha \beta} 
+ X_{\nu) \beta} X^{\beta \alpha}] 
 - \frac{1}{2} \nabla_{\alpha} \nabla_{(\mu}[ \h_{\nu) \beta} X^{\alpha \beta} 
- 2 [X] \h^{\alpha}_{\, \, \, \nu )}]    \nn \\
&&- 2 \, (M^2 + \frac{R}{2}) [ g_{\lambda\sigma}X^{~~\lambda}_{(\mu} X^\sigma_{~~\nu)} 
- [X] X_{(\mu \nu)} - \frac{1}{4} g_{\mu \nu} (X^{\alpha \beta} X_{\beta \alpha} 
- [X]^2) ] \nn\\
& & - \frac{1}{2} [  R_{\mu \nu} (X^{\alpha \beta} X_{\beta \alpha} 
- [X]^2) - \nabla_{(\mu} \nabla_{\nu)} (X^{\alpha \beta} X_{\beta \alpha} - [X]^2) + g_{\mu \nu} 
{\DAlambert} (X^{\alpha \beta} X_{\beta \alpha} - [X]^2) ] \nn.
\ee
Here $\h_{\mu\nu}=X_{\mu\nu}+X_{\nu\mu}$ and  $[X]=X^\alpha_{~\alpha}$ while 
\be
\mathcal{A}^{\, \, \, \; \; \lambda \alpha \beta \sigma \rho \tau}_{\mu \nu} &=& 
\delta^{\lambda}_{(\mu} \delta^{\rho}_{\nu)} g^{\alpha \tau} g^{\beta \sigma} 
- \frac{1}{2} \delta^{\lambda}_{(\mu} \delta^{\rho}_{\nu)} g^{\alpha \beta} g^{\sigma \tau} 
+ \frac{1}{2} \delta^{\alpha}_{\mu} \delta^{\beta}_{\nu} g^{\lambda \rho} g^{\sigma \tau} 
- \frac{1}{2} \delta^{\rho}_{\mu} \delta^{\alpha}_{\nu} g^{\lambda \tau} g^{\sigma \beta}  \nn \\
&&- \frac{1}{2} \delta^{\rho}_{\mu} \delta^{\alpha}_{\nu} g^{\lambda \sigma} g^{\tau \beta}
 + \frac{1}{4} \delta^{\rho}_{\mu} \delta^{\tau}_{\nu} g^{\alpha \beta} g^{\lambda \sigma}
  - \frac{1}{4} \delta^{\lambda}_{\mu} \delta^{\sigma}_{\nu} g^{\alpha \rho} g^{\beta \tau} 
  + \frac{1}{4} \delta^{\lambda}_{\mu} \delta^{\sigma}_{\nu} g^{\alpha \beta} g^{\rho \tau}  \nn \\
&&  - \frac{1}{4} g_{\mu \nu} [ g^{\lambda \tau} g^{\alpha \rho} g^{\beta \sigma}  
  + \frac{1}{2} g^{\alpha \beta} g^{\rho \tau} g^{\lambda \sigma} 
 - \frac{1}{2} g^{\alpha \rho} g^{\beta \tau} g^{\lambda \sigma} ], \nn \\ 
\mathcal{B}_{\mu \nu}^{\; \; \; \; \lambda \alpha \beta \sigma \rho \rho} &=& - \delta^{\alpha}_{(\mu} \delta^{\rho}_{\nu)} g^{\beta \sigma} g^{\lambda \tau} + \frac{1}{2} \delta^{\alpha}_{(\mu} \delta^{\sigma}_{\nu)} g^{\beta \lambda} g^{\rho \tau} + \frac{1}{2} \delta^{\alpha}_{\mu} \delta^{\beta}_{\nu} g^{\lambda \tau} g^{\sigma \rho} \nn \\
& & + \frac{1}{2} \delta^{\rho}_{\mu} \delta^{\tau}_{\nu} g^{\alpha \lambda} g^{\beta \sigma} - \frac{1}{4} \delta^{\alpha}_{\mu} \delta^{\beta}_{\nu} g^{\lambda \sigma} g^{\rho \tau} - \frac{1}{4} g_{\mu \nu} g^{\alpha \lambda} g^{\beta \sigma} g^{\rho \tau}. 
\ee
The invariance  
of the action under the spacetime diffeomorphisms implies that the  following relation should hold identically (off-shell):
\be
E^{\alpha \beta} ( \nabla_{\nu} X_{\alpha \beta} - \nabla_{\beta} X_{\nu \alpha} - \nabla_{\alpha} X_{\beta \nu} ) - X_{\nu \alpha} \nabla_{\beta} E^{\alpha \beta} - X_{\beta \nu} \nabla_{\alpha} E^{\alpha \beta} - \nabla^{\mu} T_{\mu \nu} = 0,
\ee
\color{black}
where $E_{\mu\nu}\equiv \Delta_{\mu\nu}+{\cal M}_{\mu\nu}$. To verify  our calculations, 
we checked  that this relation is indeed fulfilled for the $T_{\mu\nu}$ given by the above formulas.


\begin{thebibliography}{10}

\bibitem{Fierz:1939ix}
M.~Fierz and W.~Pauli, \emph{{On relativistic wave equations for particles of
  arbitrary spin in an electromagnetic field}},
  \href{http://dx.doi.org/10.1098/rspa.1939.0140}{\emph{Proc.Roy.Soc.Lond.}
  {\bf A173} (1939) 211--232}.

\bibitem{Aragone:1971kh}
C.~Aragone and S.~Deser, \emph{{Constraints on gravitationally coupled tensor
  fields}}, \href{http://dx.doi.org/10.1007/BF02813572}{\emph{Nuovo Cim.} {\bf
  A3} (1971) 709--720}.

\bibitem{Aragone:1979bm}
C.~Aragone and S.~Deser, \emph{{Consistency Problems of Spin-2 Gravity
  Coupling}}, \href{http://dx.doi.org/10.1007/BF02722400}{\emph{Nuovo Cim.}
  {\bf B57} (1980) 33--49}.

\bibitem{Higuchi}
A.~Higuchi, \emph{{Forbidden mass range for spin-2 field theory in de Sitter
  space-time}},
  \href{http://dx.doi.org/10.1016/0550-3213(87)90691-2}{\emph{Nucl.Phys.} {\bf
  B282} (1987) 397}.

\bibitem{Buchbinder:1999ar}
I.~L. Buchbinder, D.~M. Gitman, V.~A. Krykhtin and V.~D. Pershin,
  \emph{{Equations of motion for massive spin-2 field coupled to gravity}},
  \href{http://dx.doi.org/10.1016/S0550-3213(00)00389-8}{\emph{Nucl. Phys.}
  {\bf B584} (2000) 615--640},
  [\href{https://arxiv.org/abs/hep-th/9910188}{{\tt hep-th/9910188}}].

\bibitem{Bernard:2014bfa}
L.~Bernard, C.~Deffayet and M.~von Strauss, \emph{{Consistent massive graviton
  on arbitrary backgrounds}},
  \href{http://dx.doi.org/10.1103/PhysRevD.91.104013}{\emph{Phys.Rev.} {\bf
  D91} (2015) 104013}, [\href{https://arxiv.org/abs/1410.8302}{{\tt
  1410.8302}}].

\bibitem{Bernard:2015mkk}
L.~Bernard, C.~Deffayet and M.~von Strauss, \emph{{Massive graviton on
  arbitrary background: derivation, syzygies, applications}},
  \href{http://dx.doi.org/10.1088/1475-7516/2015/06/038}{\emph{JCAP} {\bf 1506}
  (2015) 038}, [\href{https://arxiv.org/abs/1504.04382}{{\tt 1504.04382}}].

\bibitem{Bernard:2015uic}
L.~Bernard, C.~Deffayet, A.~Schmidt-May and M.~von Strauss, \emph{{Linear
  spin-2 fields in most general backgrounds}},
  \href{http://dx.doi.org/10.1103/PhysRevD.93.084020}{\emph{Phys. Rev.} {\bf
  D93} (2016) 084020}, [\href{https://arxiv.org/abs/1512.03620}{{\tt
  1512.03620}}].

\bibitem{deRham:2010kj}
C.~de~Rham, G.~Gabadadze and A.~Tolley, \emph{{Resummation of massive
  gravity}},
  \href{http://dx.doi.org/10.1103/PhysRevLett.106.231101}{\emph{Phys.Rev.Lett.}
  {\bf 106} (2011) 231101}, [\href{https://arxiv.org/abs/1011.1232}{{\tt
  1011.1232}}].

\bibitem{Deser:2015wta}
S.~Deser, A.~Waldron and G.~Zahariade, \emph{{Propagation peculiarities of mean
  field massive gravity}},
  \href{http://dx.doi.org/10.1016/j.physletb.2015.07.055}{\emph{Phys. Lett.}
  {\bf B749} (2015) 144--148}, [\href{https://arxiv.org/abs/1504.02919}{{\tt
  1504.02919}}].

\bibitem{Chamseddine:2011mu}
A.~H. Chamseddine and V.~Mukhanov, \emph{{Massive Gravity Simplified: A
  Quadratic Action}},
  \href{http://dx.doi.org/10.1007/JHEP08(2011)091}{\emph{JHEP} {\bf 08} (2011)
  091}, [\href{https://arxiv.org/abs/1106.5868}{{\tt 1106.5868}}].

\bibitem{Hinterbichler:2012cn}
K.~Hinterbichler and R.~A. Rosen, \emph{{Interacting Spin-2 Fields}},
  \href{http://dx.doi.org/10.1007/JHEP07(2012)047}{\emph{JHEP} {\bf 07} (2012)
  047}, [\href{https://arxiv.org/abs/1203.5783}{{\tt 1203.5783}}].

\bibitem{Deffayet:2012zc}
C.~Deffayet, J.~Mourad and G.~Zahariade, \emph{{A note on 'symmetric' vielbeins
  in bimetric, massive, perturbative and non perturbative gravities}},
  \href{http://dx.doi.org/10.1007/JHEP03(2013)086}{\emph{JHEP} {\bf 03} (2013)
  086}, [\href{https://arxiv.org/abs/1208.4493}{{\tt 1208.4493}}].

\bibitem{Mazuet:2017hey}
C.~Mazuet and M.~S. Volkov, \emph{{Massive gravitons in arbitrary spacetimes}},
  \href{http://dx.doi.org/10.1103/PhysRevD.96.124023}{\emph{Phys. Rev.} {\bf
  D96} (2017) 124023}, [\href{https://arxiv.org/abs/1708.03554}{{\tt
  1708.03554}}].

\bibitem{Guarato:2013gba}
P.~Guarato and R.~Durrer, \emph{{Perturbations for massive gravity theories}},
  \href{http://dx.doi.org/10.1103/PhysRevD.89.084016}{\emph{Phys. Rev.} {\bf
  D89} (2014) 084016}, [\href{https://arxiv.org/abs/1309.2245}{{\tt
  1309.2245}}].

\bibitem{Bernard:2017tcg}
L.~Bernard, C.~Deffayet, K.~Hinterbichler and M.~von Strauss, \emph{{Partially
  Massless Graviton on Beyond Einstein Spacetimes}},
  \href{http://dx.doi.org/10.1103/PhysRevD.95.124036}{\emph{Phys. Rev.} {\bf
  D95} (2017) 124036}, [\href{https://arxiv.org/abs/1703.02538}{{\tt
  1703.02538}}].

\bibitem{Hassan:2011ea}
S.~Hassan and R.~A. Rosen, \emph{{Confirmation of the secondary constraint and
  absence of ghost in massive gravity and bimetric gravity}},
  \href{http://dx.doi.org/10.1007/JHEP04(2012)123}{\emph{JHEP} {\bf 1204}
  (2012) 123}, [\href{https://arxiv.org/abs/1111.2070}{{\tt 1111.2070}}].

\bibitem{Aoki:2017ffl}
K.~Aoki and S.~Mukohyama, \emph{{Massive graviton dark matter with environment
  dependent mass: A natural explanation of the dark matter-baryon ratio}},
  \href{http://dx.doi.org/10.1103/PhysRevD.96.104039}{\emph{Phys. Rev.} {\bf
  D96} (2017) 104039}, [\href{https://arxiv.org/abs/1708.01969}{{\tt
  1708.01969}}].

\bibitem{weinberg}
S.~Weinberg, \emph{Cosmology}.
\newblock Cosmology. OUP Oxford, 2008.

\bibitem{Dubovsky:2004ud}
S.~L. Dubovsky, P.~G. Tinyakov and I.~I. Tkachev, \emph{{Massive graviton as a
  testable cold dark matter candidate}},
  \href{http://dx.doi.org/10.1103/PhysRevLett.94.181102}{\emph{Phys. Rev.
  Lett.} {\bf 94} (2005) 181102},
  [\href{https://arxiv.org/abs/hep-th/0411158}{{\tt hep-th/0411158}}].

\bibitem{Aoki:2016zgp}
K.~Aoki and S.~Mukohyama, \emph{{Massive gravitons as dark matter and
  gravitational waves}},
  \href{http://dx.doi.org/10.1103/PhysRevD.94.024001}{\emph{Phys. Rev.} {\bf
  D94} (2016) 024001}, [\href{https://arxiv.org/abs/1604.06704}{{\tt
  1604.06704}}].

\bibitem{Babichev:2016bxi}
E.~Babichev, L.~Marzola, M.~Raidal, A.~Schmidt-May, F.~Urban, H.~Veermae
  et~al., \emph{{Heavy spin-2 Dark Matter}},
  \href{http://dx.doi.org/10.1088/1475-7516/2016/09/016}{\emph{JCAP} {\bf 1609}
  (2016) 016}, [\href{https://arxiv.org/abs/1607.03497}{{\tt 1607.03497}}].

\bibitem{Aoki:2017cnz}
K.~Aoki and K.-i. Maeda, \emph{{Condensate of Massive Graviton and Dark
  Matter}}, \href{http://dx.doi.org/10.1103/PhysRevD.97.044002}{\emph{Phys.
  Rev.} {\bf D97} (2018) 044002}, [\href{https://arxiv.org/abs/1707.05003}{{\tt
  1707.05003}}].

\bibitem{Abbott:2016blz}
{\scshape Virgo, LIGO Scientific} collaboration, B.~P. Abbott et~al.,
  \emph{{Observation of Gravitational Waves from a Binary Black Hole Merger}},
  \href{http://dx.doi.org/10.1103/PhysRevLett.116.061102}{\emph{Phys.Rev.Lett.}
  {\bf 116} (2016) 061102}, [\href{https://arxiv.org/abs/1602.03837}{{\tt
  1602.03837}}].

\bibitem{deRham:2016nuf}
C.~de~Rham, J.~T. Deskins, A.~J. Tolley and S.-Y. Zhou, \emph{{Graviton Mass
  Bounds}}, \href{http://dx.doi.org/10.1103/RevModPhys.89.025004}{\emph{Rev.
  Mod. Phys.} {\bf 89} (2017) 025004},
  [\href{https://arxiv.org/abs/1606.08462}{{\tt 1606.08462}}].

\bibitem{Deser:2012qx}
S.~Deser and A.~Waldron, \emph{{Acausality of Massive Gravity}},
  \href{http://dx.doi.org/10.1103/PhysRevLett.110.111101}{\emph{Phys.Rev.Lett.}
  {\bf 110} (2013) 111101}, [\href{https://arxiv.org/abs/1212.5835}{{\tt
  1212.5835}}].

\bibitem{Benini:2010pr}
F.~Benini, C.~P. Herzog, R.~Rahman and A.~Yarom, \emph{{Gauge gravity duality
  for d-wave superconductors: prospects and challenges}},
  \href{http://dx.doi.org/10.1007/JHEP11(2010)137}{\emph{JHEP} {\bf 11} (2010)
  137}, [\href{https://arxiv.org/abs/1007.1981}{{\tt 1007.1981}}].

\bibitem{Baggioli:2014roa}
M.~Baggioli and O.~Pujolas, \emph{{Electron-Phonon Interactions,
  Metal-Insulator Transitions, and Holographic Massive Gravity}},
  \href{http://dx.doi.org/10.1103/PhysRevLett.114.251602}{\emph{Phys. Rev.
  Lett.} {\bf 114} (2015) 251602}, [\href{https://arxiv.org/abs/1411.1003}{{\tt
  1411.1003}}].

\bibitem{Volkov:2013roa}
M.~Volkov, \emph{{Self-accelerating cosmologies and hairy black holes in
  ghost-free bigravity and massive gravity}},
  \href{http://dx.doi.org/10.1088/0264-9381/30/18/184009}{\emph{Class.Quant.Grav.}
  {\bf 30} (2013) 184009}, [\href{https://arxiv.org/abs/1304.0238}{{\tt
  1304.0238}}].

\bibitem{Bardeen:1972fi}
J.~M. Bardeen, W.~H. Press and S.~A. Teukolsky, \emph{{Rotating black holes:
  Locally nonrotating frames, energy extraction, and scalar synchrotron
  radiation}}, \href{http://dx.doi.org/10.1086/151796}{\emph{Astrophys. J.}
  {\bf 178} (1972) 347}.

\bibitem{Starobinsky:1973aij}
A.~A. Starobinsky, \emph{{Amplification of waves reflected from a rotating
  "black hole".}}, {\emph{Sov. Phys. JETP} {\bf 37} (1973) 28--32}.

\bibitem{Brito:2015oca}
R.~Brito, V.~Cardoso and P.~Pani, \emph{{Superradiance}},
  \href{http://dx.doi.org/10.1007/978-3-319-19000-6}{\emph{Lect. Notes Phys.}
  {\bf 906} (2015) pp.1--237}, [\href{https://arxiv.org/abs/1501.06570}{{\tt
  1501.06570}}].

\bibitem{Herdeiro:2014goa}
C.~A.~R. Herdeiro and E.~Radu, \emph{{Kerr black holes with scalar hair}},
  \href{http://dx.doi.org/10.1103/PhysRevLett.112.221101}{\emph{Phys. Rev.
  Lett.} {\bf 112} (2014) 221101}, [\href{https://arxiv.org/abs/1403.2757}{{\tt
  1403.2757}}].

\bibitem{Damour:1976kh}
T.~Damour, N.~Deruelle and R.~Ruffini, \emph{{On Quantum Resonances in
  Stationary Geometries}},
  \href{http://dx.doi.org/10.1007/BF02725534}{\emph{Lett. Nuovo Cim.} {\bf 15}
  (1976) 257--262}.

\bibitem{East:2017ovw}
W.~E. East and F.~Pretorius, \emph{{Superradiant Instability and Backreaction
  of Massive Vector Fields around Kerr Black Holes}},
  \href{http://dx.doi.org/10.1103/PhysRevLett.119.041101}{\emph{Phys. Rev.
  Lett.} {\bf 119} (2017) 041101},
  [\href{https://arxiv.org/abs/1704.04791}{{\tt 1704.04791}}].

\bibitem{Herdeiro:2016tmi}
C.~Herdeiro, E.~Radu and H.~Runarsson, \emph{{Kerr black holes with Proca
  hair}}, \href{http://dx.doi.org/10.1088/0264-9381/33/15/154001}{\emph{Class.
  Quant. Grav.} {\bf 33} (2016) 154001},
  [\href{https://arxiv.org/abs/1603.02687}{{\tt 1603.02687}}].

\end{thebibliography}

\providecommand{\href}[2]{#2}\begingroup\raggedright\endgroup

\end{document}